\documentclass[%
 reprint,
 amsmath,amssymb,
 aps,
]{revtex4-2}
\usepackage{float,graphicx}
\usepackage{dcolumn}
\usepackage{bm}
\usepackage{xcolor}
\usepackage{hyperref}
\begin{document}

\preprint{APS/123-QED}

\title{Layer reconstruction and missing link prediction of multilayer network with a \emph{maximum a posteriori} estimation}
\author{Junyao Kuang}
\thanks{Correspondence email address: kuang@ksu.edu}
\author{Caterina Scoglio}
\affiliation{Department of Electrical and Computer Engineering, Kansas State University, Manhattan, KS 66506, USA}
\date{\today}

\begin{abstract}
From social networks to biological networks, different types of interactions among the same set of nodes characterize distinct layers, which are termed multilayer networks. Within a multilayer network, some layers, confirmed through different experiments, could be structurally similar and interdependent. In this paper, we propose a maximum a posteriori based method to study and reconstruct the structure of a target layer in a multilayer network. Nodes within the target layer are characterized by vectors, which are employed to compute edge weights. Further, to detect
structurally similar layers, we propose a novel method \textcolor{blue}{for comparing networks based on the eigenvector centrality}. Using similar layers, we obtain the parameters of the conjugate prior. With this \textcolor{blue}{maximum a posteriori algorithm}, we can reconstruct the target layer and predict missing links. We test the method on two real multilayer networks, and the results show that the maximum a posteriori estimation is promising in reconstructing the target layer even when a large number of links is missing.
\end{abstract}

\maketitle
\section{Introduction}
Network science has been widely used in different areas, such as information diffusion, infectious disease spread, and gene co-expression analysis. Through network analysis, one can study the relations among nodes, and the robustness of a system \cite{Qihui1, Qihui2}. For example, epidemiologists can predict the number of people infected by COVID-19 through epidemic analysis. Then, they can provide advice to policymakers at an early stage to curb the spreading of the disease \cite{bara}. By constructing co-expression networks, biologists can discover crucial genes (nodes) by simply choosing genes with high degree centralities, closeness centralities, or eigenvector centralities. In a typical protein network, five to seven layers are considered to represent different types of molecular interactions \cite{bianconi, kivela}, including proteolysis, genetic interaction, co-expression, etc. Such multilayer networks are obtained through biological experiments, which could be expensive and time-consuming. A layer can be particularly important but also incomplete, with many missing links. A critical research goal is to reconstruct this important layer, called the target layer, without performing the expensive experiments, but by exploiting all information embedded in the other layers. In other words, we can estimate the target layer through existing layers \cite{spur, bian2, bacco, hg1, hg2}. After constructing the target layer, researchers can devote restricted resources to the detection of edges with high probabilities. 

Various methods have been proposed to reconstruct networks and predict missing links, and most of them are based on generative models \cite{spur, bian2, bacco, ab, di, vm, newman1, newman2, newman3, peixoto1, peixoto2, newman4, newman5, vm2, peixoto3,entropy1,entropy2,entropy3}. A generative model reconstructs the network topology by fitting a stochastic network model \cite{bacco, newman1, newman2, peixoto4}, and uses a maximum-likelihood estimation (MLE) algorithm to find the optimal parameters that can describe the network. In \cite{newman1, newman2}, the authors presented a degree-correlated stochastic block model to reconstruct a single layer network. Edges are computed through the tensor product of node vectors, and the entries of the node vectors are the degrees of the nodes in each community. Authors in paper \cite{bacco} extended the single layer stochastic block model to multilayer networks and used it to predict missing links and detect overlapping communities. The authors tried all the layer combinations to find the layers that can improve the maximum likelihood. However, when there are many layers, it is burdensome to try all the layer combinations to find the interdependent layers. The authors validated the algorithm through two real multilayer networks by hiding 20\% of links and non-links. 

In multilayer networks, there could be structurally similar layers. Therefore, it is possible to take advantage of similar layers to help reconstruct the target layer. We propose comparing the target layer with the remaining layers if the target layer is partially known. In the literature, multiple methods have been proposed to compare networks \cite{corr, deltacon, compare, webgraph, simest}. The authors in \cite{deltacon, compare} present a method called DeltaCon. The DeltaCon method compares the affinity scores of every pair of nodes in two networks. The method is very sensitive to changes in the number of edges, and the removal of edges results in a significant change in the distance. Papers \cite{webgraph, simest} review and compare some network-comparing methods, including vertex/edge overlapping, vertex/edge vector similarity, and the SimHash algorithm. The vertex/edge overlapping method applies the rule that two graphs are similar if they share many vertices and edges. According to the analysis in \cite{webgraph}, the drawbacks of this method are that it is not sensitive to changes in high-quality vertices, topology, and properties of networks. The vertex/edge vector similarity method compares the node/edge weight vectors of two networks. The drawback of this method is that it is not sensitive to changes in the topology and other properties of networks. To take advantage of the features of networks, the SimHash algorithm is introduced to compare networks. The PageRank \cite{page} together with edges are used as network features in SimHash algorithm to compare web page networks.

In this paper, we propose a \emph{maximum a posteriori} (MAP) based-method for target layer reconstruction as well as for link prediction. The MLE algorithm and entropy-related approaches must depend on the known information of the target layer. Consequently, the reconstruction is significantly affected by the available information of the target layer. In the MAP algorithm, the layers that are similar to the target layer will be considered to compute the parameters of the conjugate prior. Experimental results show that the MAP algorithm provides more consistent results than the MLE method. The first contribution of this paper is to discover an incomplete target layer by computing its edges through a dot product of node vectors. The optimal entries of node vectors are obtained by maximizing the posterior probabilities of the stochastic model. In our experiments, we find that the model accuracy can be improved if we increase the dimension of node vectors, but the return is diminishing for large vector dimensions. Another contribution is that we introduce the eigenvector centrality-based SimHash algorithm to detect structurally similar layers (interdependent layers). The eigenvector centralities of nodes are extracted as network features, which allow us to recover the structure of networks, as shown in the experimental results. In this work, we assume that the number of edges between any pair of nodes follows Poisson distribution \cite{bacco, newman1}. Hence, the Gamma distribution will be the conjugate prior for the Poisson distribution \cite{map}. We compute the parameters of the conjugate prior through the adjacency matrices of similar layers, and the contributions of the similar layers are weighted by their similarities. The number of edges between each pair of nodes is calculated as the dot product of the node vectors. In our experiments, we show that similar layers are critical in improving the robustness of link predictions.

The paper is organized as follows. In section two, we first introduce the MAP method on target layer reconstruction. Next, we propose the eigenvector centrality-based SimHash algorithm to find structurally similar layers. Then, we propose the process for identifying parameters of the conjugate prior under different circumstances. In section three, we first evaluate the eigenvector centrality based SimHash algorithm on two real multilayer networks. Then, we evaluate the MAP algorithm-based target layer reconstruction on the two real networks and compare the differences between the MLE algorithm and the MAP algorithm. We conclude the paper in section four.

\section{layer reconstruction in multilayer networks} 
\subsection{\emph{Maximum a posteriori} based stochastic model}
In this section, we define the stochastic model for both directed and undirected multilayer networks. The adjacency matrix of the target layer is denoted by $A$. The goal of this reconstruction is to estimate $A$, given a partial knowledge of the target layer and of other layers in the multilayer network. To reconstruct the target layer, a set of parameters is needed to describe the model, which we denote as $\theta$. Based on Bayes’ theorem, the posterior probability of $\theta$ is
\begin{align}
\label{eq:1}
    P(\theta\ |\ A)&=\frac{P(A\ |\ \theta) P(\theta)}{P(A)},
\end{align}
where $P(\theta\ |\ A )$ is the posterior probability of $\theta$, $P(\ A\ |\ \theta)$ is the likelihood of $A$ under $\theta$, $P(\theta)$ is the prior probability of $\theta$ and $P(A)$ is the marginal likelihood that contains all the information of the network. Since $P(A)$ is a constant, $P(\theta\ |\ A)$ is proportional to the product of $P(A\ |\ \theta)$ and $P(\theta)$. Therefore, we have
\begin{align}
\label{eq:2}
    P(\theta\ |\ A) &\propto P(A\ |\ \theta) P(\theta).
\end{align}

For any pair of nodes in the network, we use $E_{ij}$ to denote the expected number of links (which could be fractional) between node $i$ and node $j$. In unweighted networks, the entries of the adjacency matrix are denoted by 0 or 1. Here, since the entries $E_{ij}$ are real numbers, we can interpret network $A$ as a weighted network.

Before we substitute any parameters into expression \eqref{eq:2}, we make the following assumptions. The links in the target layer are independent and identically distributed. In other words, the number of edges between node $i$ and node $j$ does not affect the relation between node $i$ and node $k$. Further, we assume the number of links between any pair of nodes is extracted from a Poisson distribution, i.e., $P(A_{ij}\ |\ E_{ij})=\frac{e^{-E_{ij}} (E_{ij})^{A_{ij}}}{A_{ij}!}$. We can rewrite expression \eqref{eq:2} after substituting $E_{ij}$ and $A_{ij}$ as
\begin{align}
\label{eq:3}
    P(\theta\ |\ A) &\propto \prod_{i,j}\frac{e^{-E_{ij}} (E_{ij})^{A_{ij}}}{A_{ij}!}P(E_{ij}).
\end{align}

In the MLE algorithm, the prior probability $P(E_{ij})$ can be neglected since it is a constant. In the MAP algorithm, we need to specify the prior distribution of $P(E_{ij})$. The conjugate prior distribution for the Poisson distribution is the Gamma distribution
\begin{align}
\label{eq:4}
    P(E_{ij}) &=\frac{\beta_{ij}^{\alpha_{ij}}}{\Gamma(\alpha_{ij})} E_{ij}^{\alpha_{ij}-1} e^{-\beta_{ij} E_{ij}} \nonumber\\
    &\propto E_{ij}^{\alpha_{ij}-1} e^{-\beta_{ij}  E_{ij}},
\end{align}
where $\alpha_{ij}$, $\beta_{ij}$ and $\Gamma(\alpha_{ij})$ are the shape parameter, the scale parameter, and the Gamma function of the Gamma distribution, respectively. In section II C, we introduce a procedure to determine $\alpha_{ij}$ and $\beta_{ij}$ through the layers with high similarities. Substituting the conjugate Gamma distribution into expression \eqref{eq:3}, we have
\begin{align}
\label{eq:5}
    P(\theta\ |\ A) &\propto \prod_{i,j}{e^{-(\beta+1) E_{ij}} (E_{ij})^{A_{ij}+\alpha-1}}.
\end{align}
Note that we have left out constant terms.

The problem now has been simplified to finding the parameters {$E_{ij}$} that can maximize the posterior probability. However, an expression for $E_{ij}$ has not been specified. In this work, we compute the links through node vectors. The nodes in the target layer are represented by vectors. The expected number of links $E_{ij}$ can be computed by
\begin{align}
\label{eq:6}
    E_{ij}&=\sum_{z=1}^{K} s_{iz} t_{jz},
\end{align}
where $s_{iz}$ and $t_{jz}$ are respectively the $z$th entry of node $i$'s vector and node $j$'s vector. Here, we use $s$ and $t$ to denote source and target nodes. $K$ is the dimension of the vector. Some MLE related works use tensor factorization to decompose the links, and the dimension of the tensor is interpreted as the number of overlapping communities. As a result, there should be an optimal number of communities that can maximize the estimation accuracy. However, this cannot be rigorously achieved \textcolor{blue}{and the tensor factorization can be simplified to the dot product according to our analysis in appendix A.} 

Expression \eqref{eq:5} is still intractable after substituting $E_{ij}$ with equation \eqref{eq:6}. We take the logarithmic form of expression \eqref{eq:5}, which gives
\begin{align}
\label{eq:7}
    L(\theta\ |\ A) =& \sum_{i,j}[(A_{ij}+\alpha_{ij}-1) \log E_{ij}-(\beta_{ij}+1)E_{ij}]\nonumber\\
=& \sum_{i,j}[(A_{ij}+\alpha_{ij}-1) \log \sum_z^{K} s_{iz} t_{jz} \nonumber\\
&-(\beta_{ij}+1)\sum_z^{K} s_{iz} t_{jz}],
\end{align}
where $L(\theta\ |\ A)$ is the log posterior.

To find the maximized posterior for expression \eqref{eq:7}, we apply the Jensen's inequality $\log \overline{x} \geq \overline{\log x}$, which gives
\begin{align}
\label{eq:8}
\log \sum_z^{K}s_{iz} t_{jz} &=\log \sum_z^{K}q_{ijz} \frac{s_{iz} t_{jz} }{q_{ijz}}\nonumber\\
&\geq \sum_z^{K}q_{ijz} \log  \frac{s_{iz} t_{jz}}{q_{ijz}}\nonumber\\
&=\sum_z^{K}q_{ijz} (\log s_{iz} t_{jz} -\log q_{ijz}).
\end{align}
The equality is satisfied when 
\begin{align}
\label{eq:9}
q_{ijz}=\frac{s_{iz} t_{jz}}{\sum_z  s_{iz} t_{jz}}.
    \end{align}

After substituting expression \eqref{eq:8} and equation \eqref{eq:9} into equation \eqref{eq:7}, equation \eqref{eq:7} can be simplified to
\begin{align}
\label{eq:10}
    L(\theta\ |\ D) = & \sum_{i,j,z}[(A_{ij}+\alpha_{ij}-1) q_{ijz}\log s_{iz} t_{jz} \nonumber\\
    &-(\beta_{ij}+1)s_{iz} t_{jz}].
\end{align}

Taking the derivative of equation \eqref{eq:10} and equating to zero, we obtain the values of $s_{iz}$ and $t_{iz}$, the optimal node vectors that maximize the posterior probability:
\begin{align}
\label{eq:11}
s_{iz} =\frac{\sum_j (A_{ij}+\alpha_{ij}-1) q_{ijz}}{\sum_j (\beta_{ij}+1)  t_{jz}},
\end{align}
\begin{align}
\label{eq:12}
t_{jz} =\frac{\sum_i (A_{ij}+\alpha_{ij}-1) q_{ijz}}{\sum_i (\beta_{ij}+1) s_{iz}}.
\end{align}

The procedure to obtain the optimized $s_{iz}$ and $t_{jz}$ is to assign random initial values for $s_{iz}$ and $t_{jz}$, then update equation \eqref{eq:9}, \eqref{eq:11}, and \eqref{eq:12} iteratively until equation \eqref{eq:7} converges. However, before applying the above iteration, we need to identify $\alpha_{ij}$ and $\beta_{ij}$, which are introduced in section II B and section II C.

\subsection{Similarity and layer comparison}
In section II A, we have detailed the procedure to compute the optimal node vectors by maximizing the posterior probability. The parameters $\alpha_{ij}$ and $\beta_{ij}$ for the prior distribution are required to perform the posterior probability maximization. We propose to compute $\alpha_{ij}$ and $\beta_{ij}$ through the layers of the multilayer network that are similar to the target layer. Keep in mind that there are missing links in the target layer, and the percentage of missing links is not known at all. Therefore, the primary factor for an effective network-comparing method is that the method must not be significantly affected by the percentage of missing links.
 
Networks can be characterized by multiple types of centralities, such as, degree centrality, eigenvector centrality, closeness centrality. The degree centrality measures the importance of a node by capturing the number of links the node has, while the eigenvector centrality can be regarded as an extension of degree centrality in which node's importance is also affected by its neighbors' importance. The closeness centrality of a node is the average length of shortest paths between the node and all other nodes. One or more of the centralities can describe the features of a network. To compare the target layer with the other layers, we can compare the features of the layers. For our purpose, the eigenvector centrality is selected as the network feature, and is used in the SimHash algorithm to compute similarities.

The SimHash algorithm works as follows \cite{webgraph, simest}. The feature of a network can be expressed as a set of token-value pairs $\{(v_i : w_i)\}$, where $v_i$ is a node and $w_i$ is its measure under the feature, for example its eigenvector centrality. Note that the target layer and the layer to be compared have the same set of tokens. In cryptography, any messages can be encrypted to a unique binary number (digest). Similarly, we can represent each token with a unique binary number with $\phi$ bits ($2^\phi >$ the number of $v_i$). For each binary number (digest), we map every 1 to $w_i$, and 0 to $-w_i$. Thus, each token is mapped to a weighted digest with $\phi$ digits. To obtain the weighted digest of the network, we sum up all the weighted digests of the tokens. Note that there is no carry in the summation.

To measure the similarities between the target layer and the other layers, we can compare the digests of the layers. A simple way to compare the digests is to convert the weighted digests to binary digests. The binary digest of a network can be obtained by setting positive digits to 1 and negative digits to 0. The similarity between the target layer $m$ and any other layer $r$ can be measured by
 \begin{align}
\label{eq:13}
\mu_{m,r}=1-\frac{Hamming(H_m^d,\ H_r^d)}{\phi},
\end{align}
where $\mu_{m,r}$ is the similarity between layer $m$ and $r$, $H_m^d$ is the binary digest of the target layer, and $H_r^d$ is the binary digest of the layer to be compared, respectively.

An alternative way to obtain the similarities is by computing the Pearson correlation coefficient of the weighted digests.

\textcolor{blue}{The estimation results based on the Pearson correlation are shown in appendix B. In this work,} we measure the similarity \textcolor{blue}{between} the target layer and the other layers through the binary-based digests. \textcolor{blue}{The influence of the bit number is discussed in appendix C.}

\subsection{Identify the parameters of the Gamma distribution prior}
Finally, we introduce the procedure to determine the parameters $\alpha_{ij}$ and $\beta_{ij}$ of the conjugate prior. We discuss this problem in two cases.

In the first case, we assume the structure of the target layer is partially known, i.e., the entries of the adjacency matrix are partially known. In this case, we apply the layer comparison method introduced above, and the $L^\prime$ layers with highest similarities are considered to identify the parameters of conjugate priors. The parameters of the Gamma distribution prior are computed as
\begin{align}
\label{eq:14}
\begin {cases}
		\alpha_{ij} =max(\sum_{r = 1}^{L^\prime} \mu_{m, r} A_{ij}^r,\ 1)  \\
        \beta_{ij} =max(\sum_{r = 1}^{L^\prime} \mu_{m, r},\ \frac{\sum_{r =1}^{L^\prime} \mu_{m, r}}{\sum_{r = 1}^{L^\prime} \mu_{m, r} A_{ij}^r}), 
\end{cases}
\end{align}
where $m$ denotes the target layer, $\mu_{m, r}$ is the similarity between the target layer $m$ and any layer $r$. $A_{ij}^r$ is the adjacency matrix of layer $r$ in $L^\prime$. In equations \eqref{eq:11} and \eqref{eq:12}, since $A_{ij}$ could be zero, if $\alpha_{ij}$ is less than one, we obtain a negative $s_{ij}$. Thus, we need to limit the range of $\alpha_{ij}$. If $\sum_{r=1}^{L^\prime} \mu_{m, r} A_{ij}^r \in (0, 1)$ , we set $\alpha_{ij} =1$, and set $\beta_{ij}$ to $\frac{\sum_{r=1}^{L^\prime} \mu_{m, r}}{\sum_{r = 1}^{L^\prime} \mu_{m, r} A_{ij}^r}$ to maintain the means of the Gamma distribution prior unchanged. If $\sum_{r=1}^{L^\prime} \mu_{m, r} A_{ij}^r =0$, we will set $\alpha_{ij}=1$, and set $\beta_{ij}$ equal to a large number to ensure the MAP algorithm converges. 

A special case is the structure of the target layer is not known at all. In this case, the comparison between the target layer and the other layers is not feasible. In this case, an alternative and heuristic way to compute the parameters of conjugate prior can be based on the functionally similar layers. If available, we can use additional published networks and data as a new multilayer network, in which layers are functionally similar. Thereafter, we can apply the proposed MAP algorithm to compute the parameters of conjugate prior through this new multilayer network. The similarities between the target layer and the functionally layers cannot be determined through any network comparison algorithm. Thus, we assume the similarities are all ones. We assign the parameters of the Gamma distribution as
\begin{align}
\label{eq:15}
\begin {cases}
		\alpha_{ij} =\sum_{r=1}^{L^\prime} A_{ij}^r \\
        \beta_{ij} =L^\prime.
\end{cases}
\end{align}
Similarly, if $\sum_{r =1}^L A_{ij}^r<1$, we will set $\alpha_{ij}=1$, and set $\beta_{ij}$ to a large number to make the MAP algorithm converge.

In this scenario, we do not have any structural information regarding the target layer. If we set $A_{ij}=0$ in equations \eqref{eq:11} and \eqref{eq:12}, this equivalent to assuming zero presences of all the edges in the target layer. To avoid this issue, we take the entries of $A_{ij}$ as the ratio of $\alpha_{ij}$ and $\beta_{ij}$, i.e., $A_{ij}=\alpha_{ij}/\beta_{ij}$. The entries are assigned as the average presences over the other layers.

\section{experimental validation}
In this section, we evaluate the method introduced in section II. First, the SimHash algorithm is applied on two real networks, in which different percentages of links are removed uniformly at random. Second, the effectiveness of the MAP algorithm for the two real multilayer networks is evaluated, and the MAP algorithm is compared to the MLE algorithm.

\begin{figure}[ht]
\includegraphics[width=0.48\linewidth]{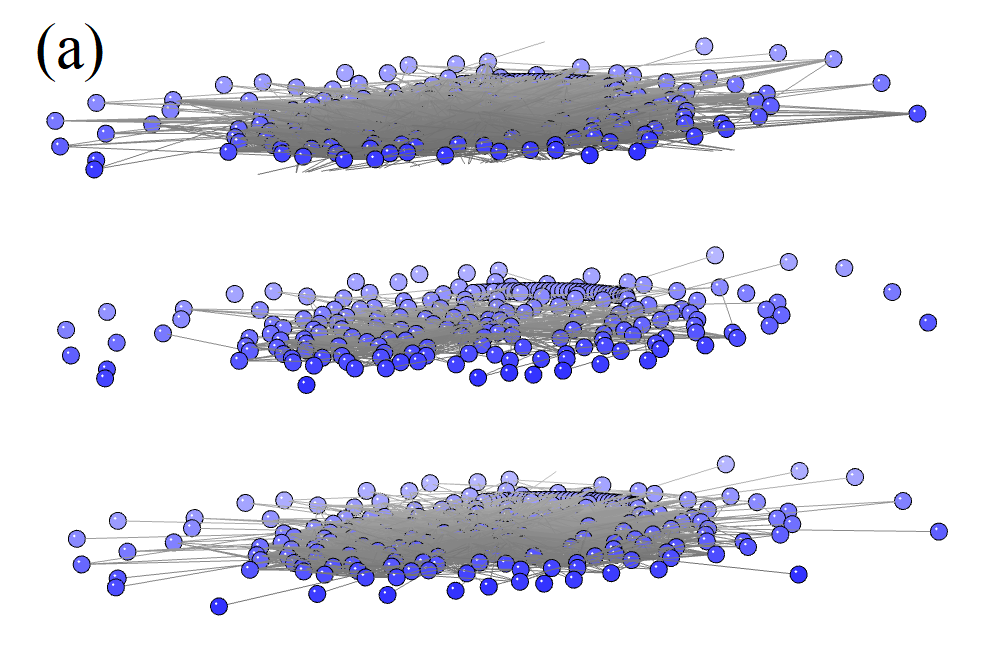}
\includegraphics[width=0.48\linewidth]{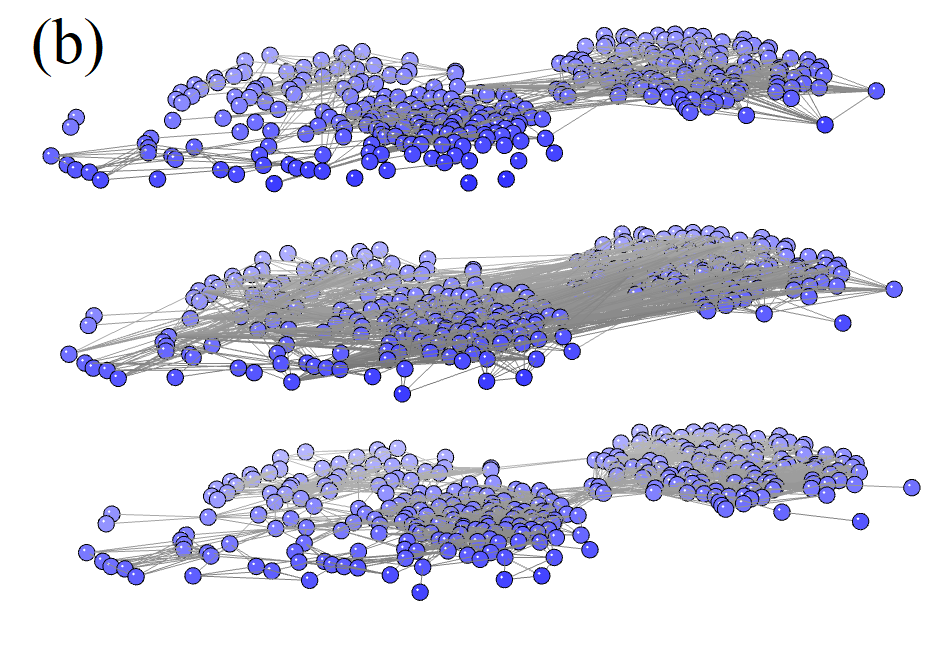}
\caption{\label{fig:fig1} Three layers of the FAO \textcolor{blue}{(a)} network and HVR \textcolor{blue}{(b)} network. Nodes in different layers share the same plane coordinates.}
\end{figure}

The first real network we use to evaluate our proposed MAP algorithm is the FAO (Food and Agriculture Organization) trade network \cite{fao}. The FAO multilayer network is composed of 364 layers, and each layer represents a product trading among 214 countries. A link is detected between two nodes in a layer if there is trading of the corresponding product between the two countries. We show three layers of the FAO network in Fig. \ref{fig:fig1} through the KiNG software \cite{king}. Since the layers in the FAO network are not ordered in any particular way, in the experiments, we only perform the evaluation by assuming that the target layer is one of the the first nine layers of the FAO network. Note that all the 363 remaining layers are always used in our experiments for detecting the similar layers.

The second real network (HVR network) we use to evaluate the proposed MAP algorithm has nine layers and 307 nodes \cite{hvr}, which represent malaria parasite genes. The nine layers correspond to nine highly variable regions on the genes themselves. An edge is detected if two genes share an exact match of significant length in the highly variable region. We show three layers of the HVR network in Fig. \ref{fig:fig1}.

\subsection{Numerical results on layer comparison}

\begin{figure}[ht]
\includegraphics[width=0.98\linewidth]{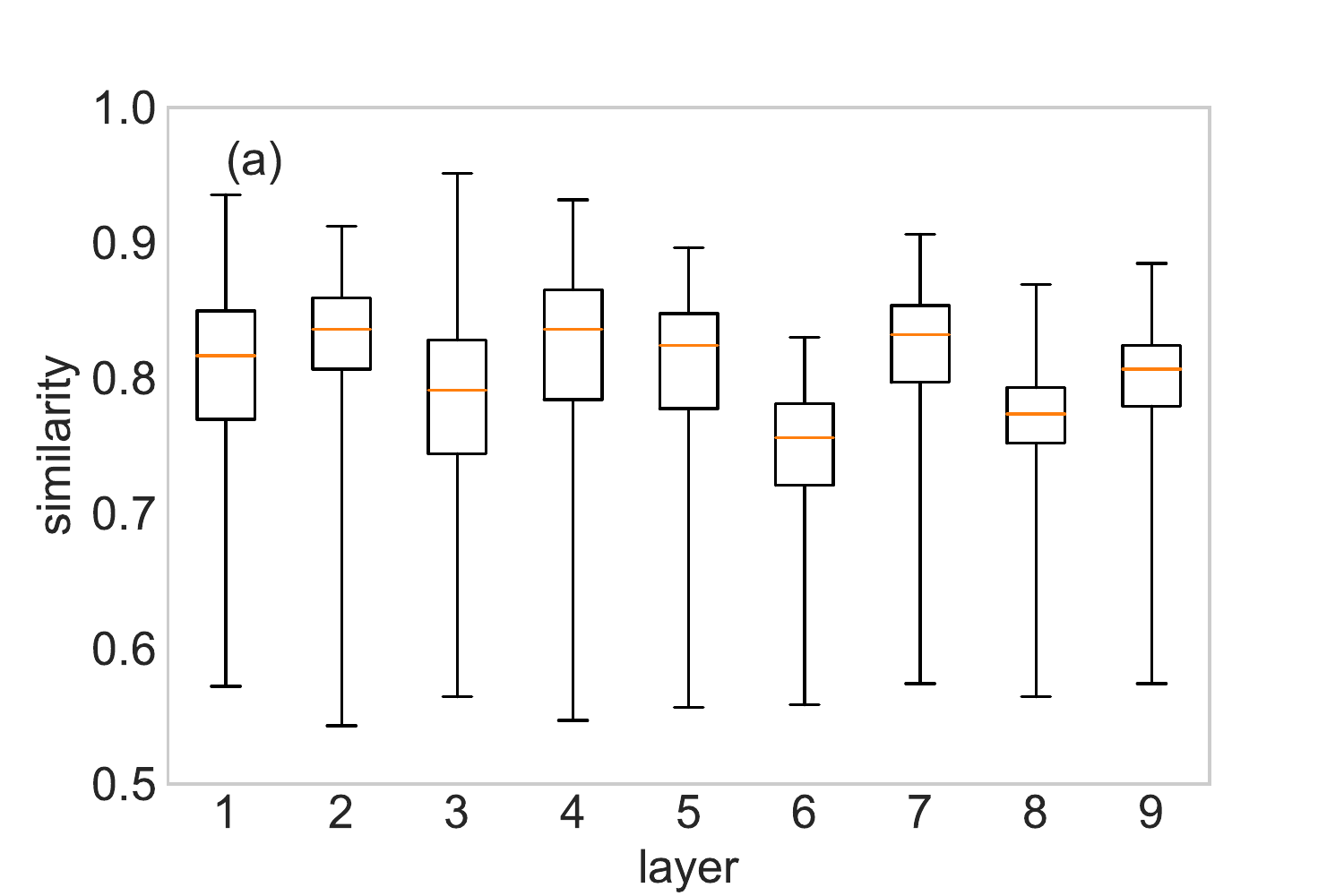}
\includegraphics[width=0.98\linewidth]{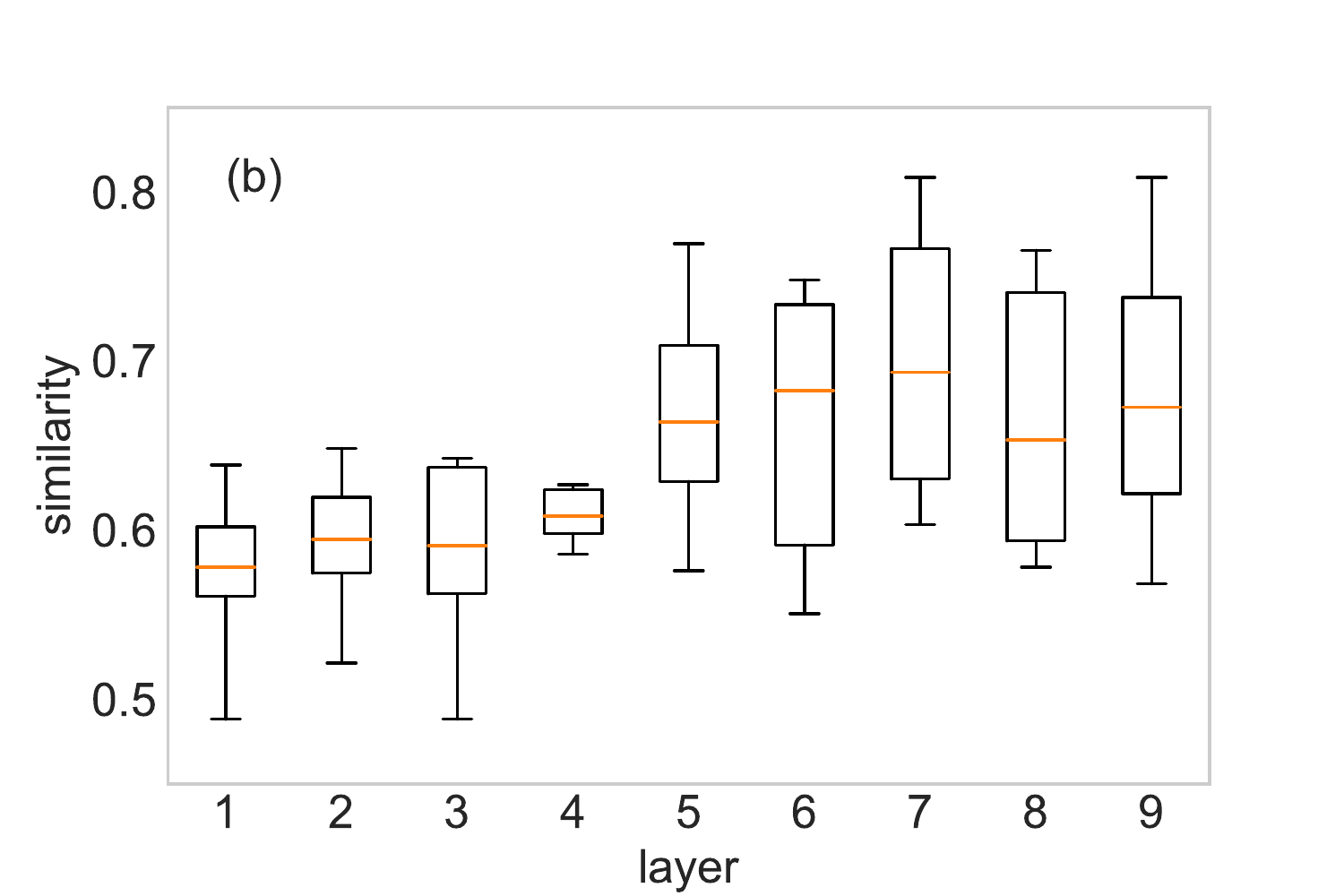}
\caption{\label{fig:fig2} Layer comparison of the FAO network \textcolor{blue}{(a)} and HVR network \textcolor{blue}{(b). (a) shows} the similarities between the target layer and all the other 363 layers in the FAO network. \textcolor{blue}{(b) shows} the similarities between the target layer and the other eight layers in the HVR network. Note that each similarity value is averaged over 10 runs.}
\end{figure}

The evaluation of the SimHash algorithm is performed on both the FAO network and the HVR network. The binary digest of each layer is obtained through the SimHash algorithm based on the eigenvetor centrality of the nodes. The similarity of any two layers can be computed through equation \eqref{eq:13}. In Fig. \ref{fig:fig2}, each layer is respectively set as the target layer, and the similarities between the target layer and all the other layers are shown in interquartile ranges (IQR). For the FAO network, we only show the results of the first nine layers as target layers, and each IQR bar contains 363 similarity values. The similarities obtained through the SimHash algorithm are between 0 and 1. A zero similarity means the two layers have totally opposite eigenvector centrality distribution, while for 0.5 similarity the two layers are independent. A similarity approaching one means the two layers are similar. In the FAO network, the similarities are between 0.5 and 1, which indicates there are no layers with totally opposite eigenvector centrality distribution. In the HVR network, most of the layers are independent, since the similarities are all less than 0.8, except layer 7 and layer 9. 

\begin{figure}[ht]
\includegraphics[width=0.48\linewidth]{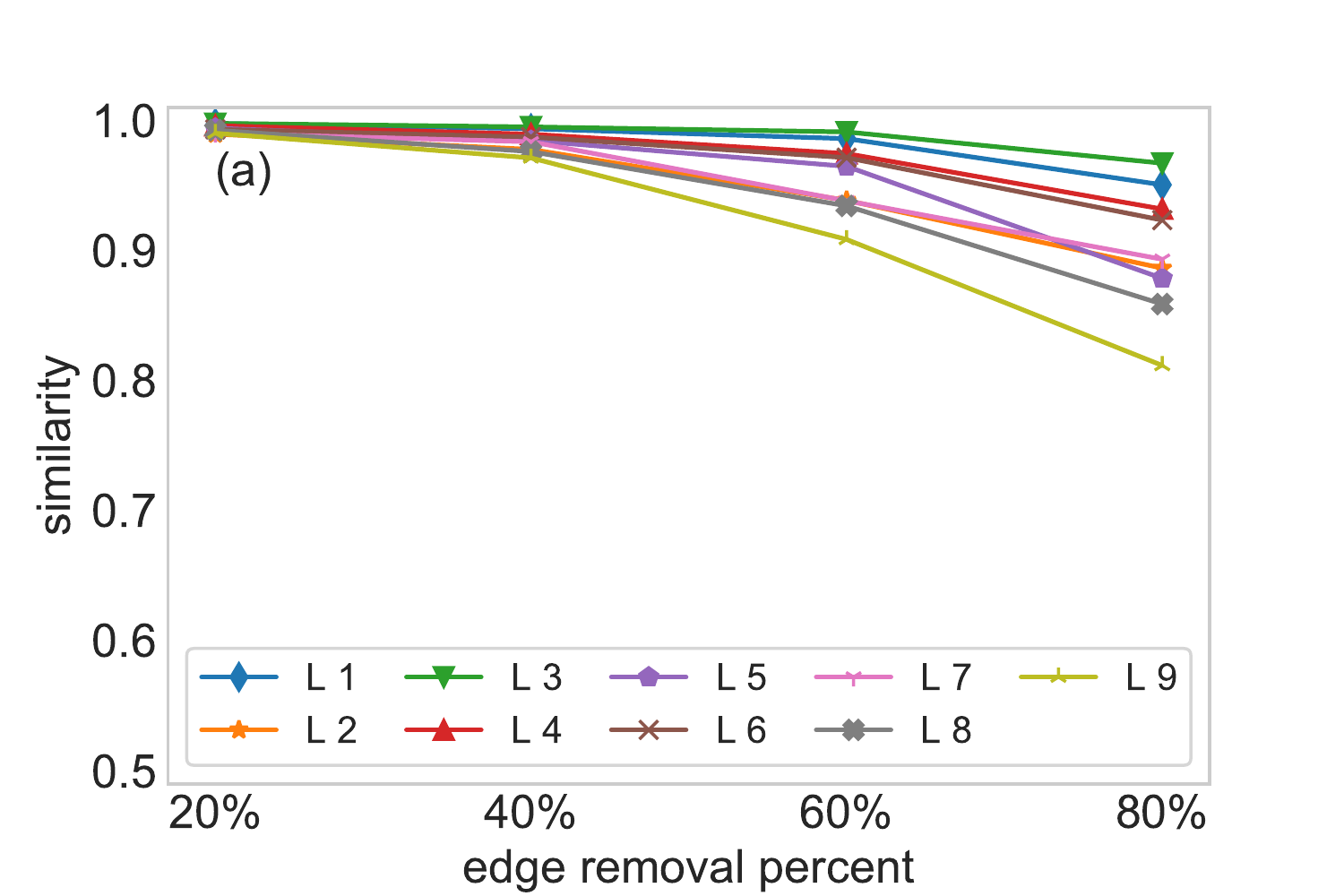}
\includegraphics[width=0.48\linewidth]{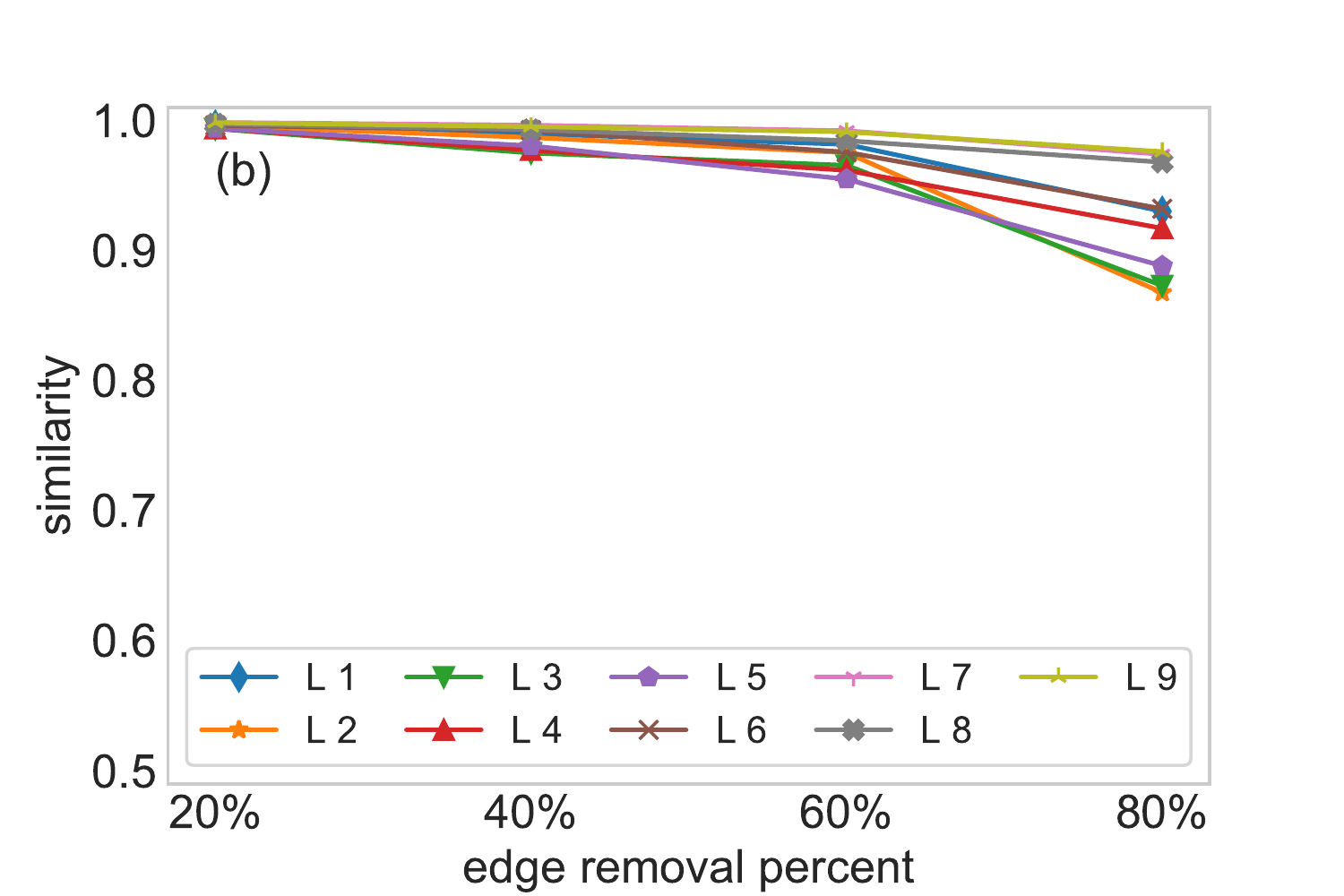}
\includegraphics[width=0.48\linewidth]{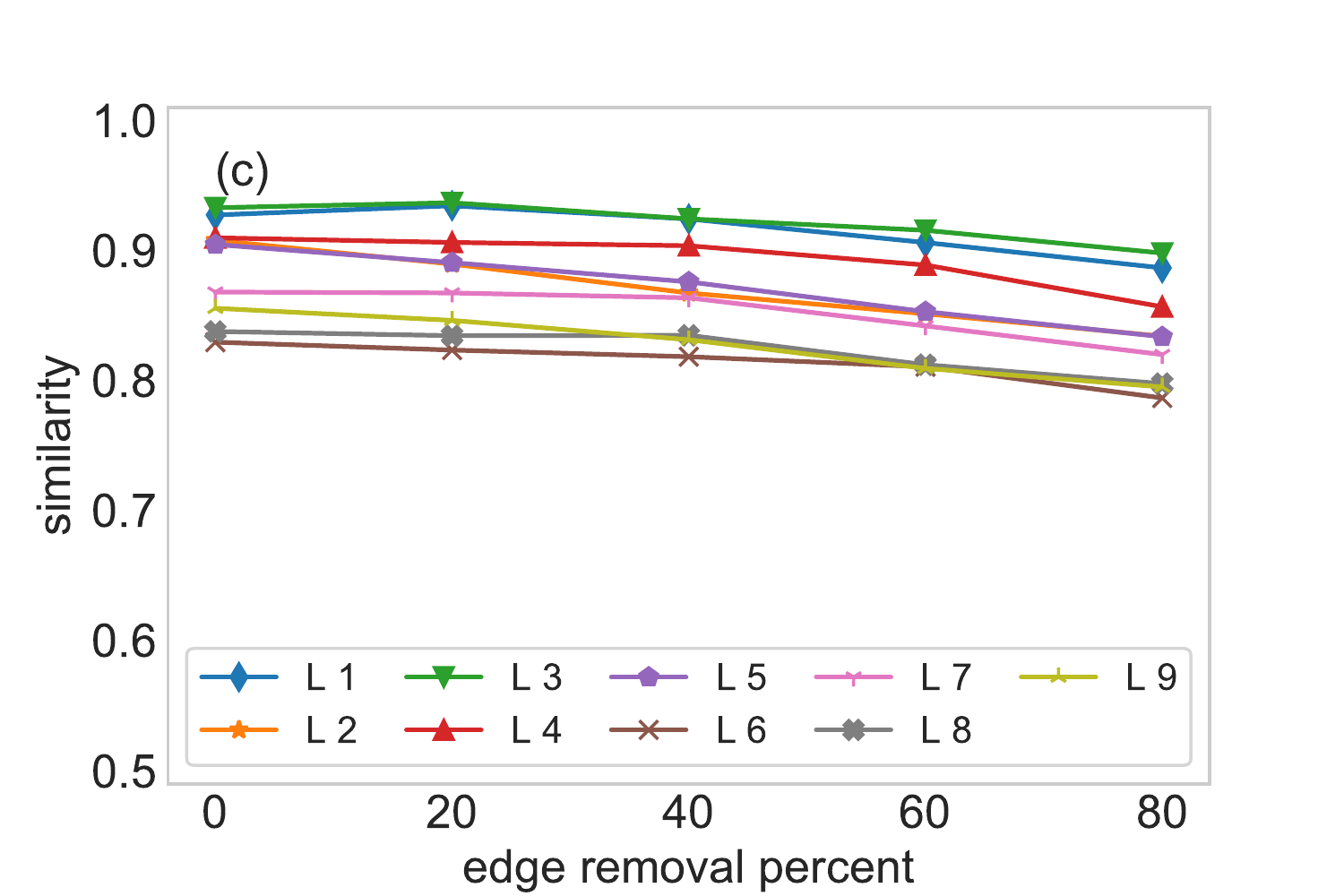}
\includegraphics[width=0.48\linewidth]{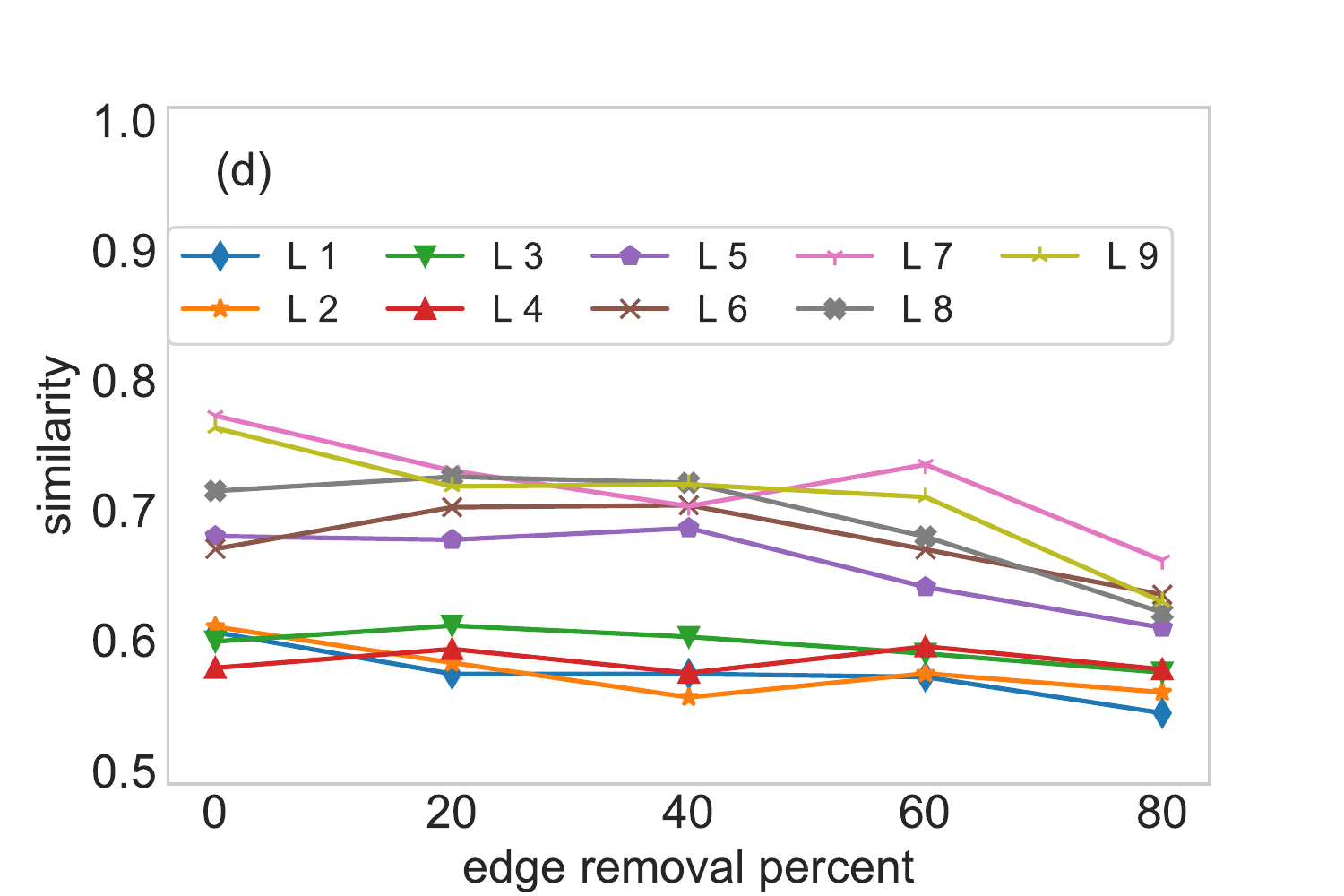}
\caption{\label{fig:fig3} Eigenvector centrality-based SimHash algorithm. \textcolor{blue}{Panel (a)} shows the comparison between the first nine layers of the FAO network and their incomplete counterparts, where 20\%, 40\%, 60\% and 80\% edges are removed uniformly at random from the target layer. \textcolor{blue}{Panel (b)} shows the same experiment on the HVR network. \textcolor{blue}{Panel (c)} compares the reduced networks (the first nine layers of the FAO network) with the five most similar layers. The mean of the similarities is shown in the panel. Similarly, 20\%, 40\%, 60\% and 80\% edges are removed uniformly at random from the target layer. \textcolor{blue}{Panel (d)} shows the same experiment on the HVR network. Note that each similarity value is averaged over 10 runs.}
\end{figure}

The network-comparing method must be able to discover structurally similar layers even if there are missing links in the network. To this end, we uniformly at random remove 20\%, 40\%, 60\% and 80\% of edges from the target layer to generate incomplete networks and compare the incomplete networks with the target network. In \textcolor{blue}{Fig. \ref{fig:fig3}(a)}, each layer of the FAO network (first nine layers) and HVR network is set as target network respectively, and we randomly remove 20\%, 40\%, 60\% and 80\% of edges from the target layer. The similarities between the incomplete networks with removed edges and the target layer are computed. From the top panels, we observe that the similarities are greater than 0.8 even with 80\% of edges randomly removed.

There are always differences between the target layer and similar layers. In the second experiment, we show that missing links in the target layers do not affect the similarity significantly between the target layer and its similar layers. For each of the target layers, we choose five most similar layers from all the other layers, we then randomly remove 20\%, 40\%, 60\% and 80\% of edges from the target layer to generate incomplete networks. The incomplete networks are compared to the five similar layers, and we can obtain five similarities for different removal percentages. In the bottom panels, we show the average of the five similarities. The results for the first nine layers of the FAO network are shown in \textcolor{blue}{Fig. \ref{fig:fig3}(c)}. We observe that the similarities decrease slightly even when 80\% of edges are randomly removed. For the HVR network, where the layers are heterogeneous, though we randomly remove different percentages of links, the similarities are still maintained at low levels.

\subsection{Validation of layer reconstruction and link prediction}
The MAP method we have introduced in this work has the goal of layer reconstruction and missing link estimation. The similarity between the target layer and other layers can be obtained through the SimHash algorithm as introduced above. In this part, we show the effectiveness of the MAP algorithm, and compare the differences between the MAP algorithm and the MLE algorithm performance. Other methods such as entropy-based approaches, are equivalent to the MLE algorithm and are discussed \textcolor{blue}{in appendix D}.

Here, we use the receiver-operator characteristic (ROC) curve and the area under the curve (AUC) to evaluate the effectiveness of our method. The model is perfect when the AUC is approaching one, and 0.5 means the model guesses the edge weights randomly.

\begin{figure}[h]
\includegraphics[width=0.48\linewidth]{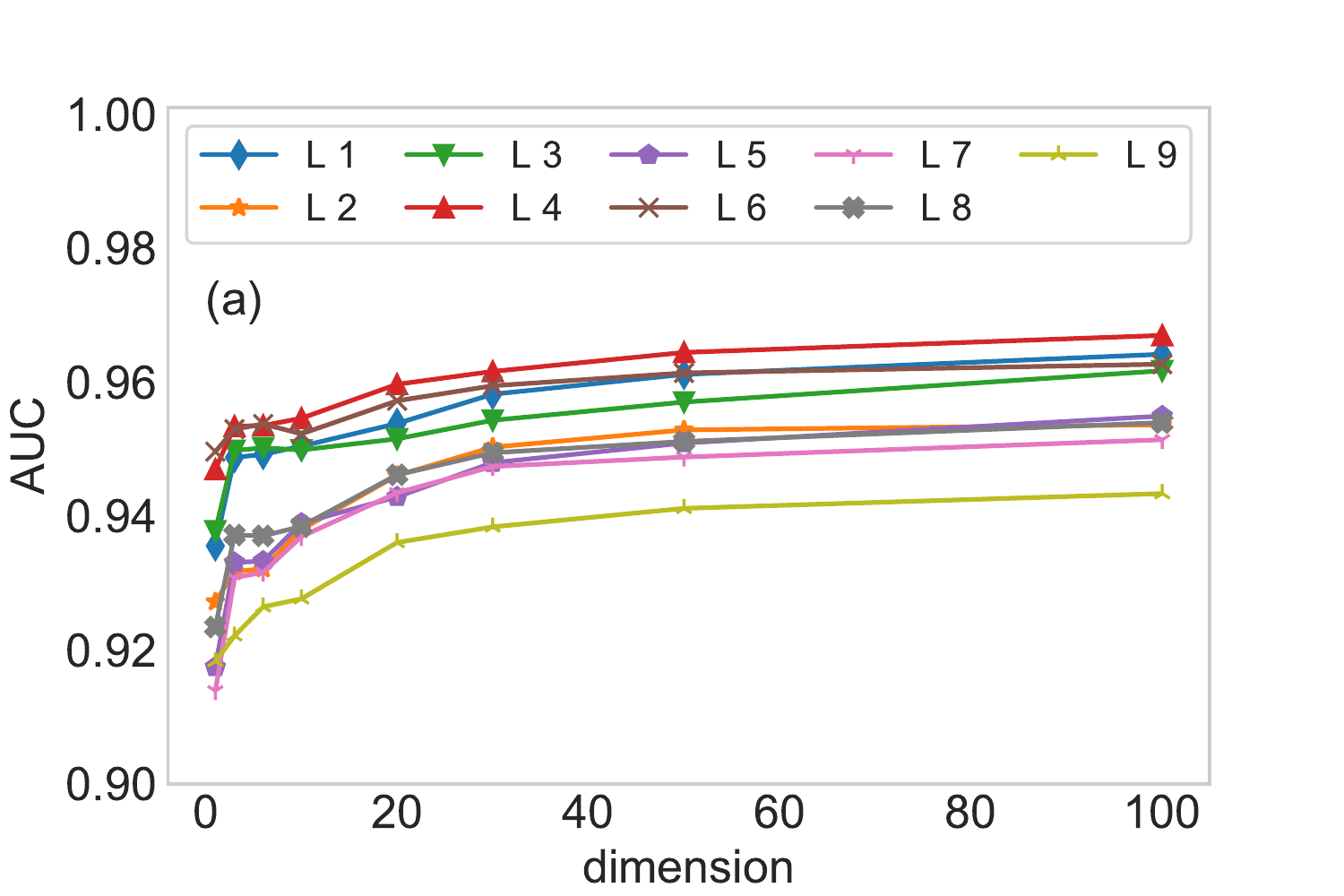}
\includegraphics[width=0.48\linewidth]{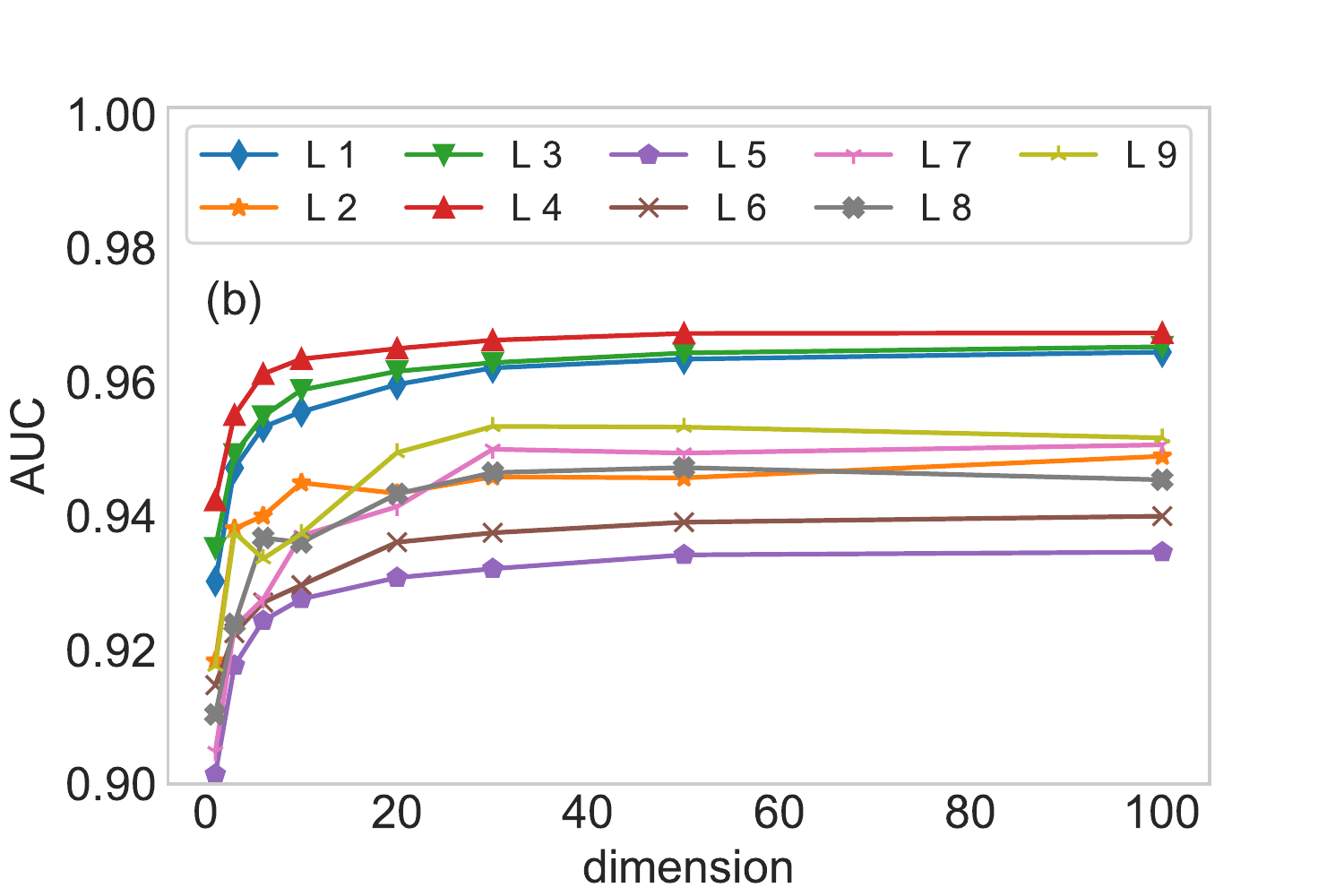}
\includegraphics[width=0.48\linewidth]{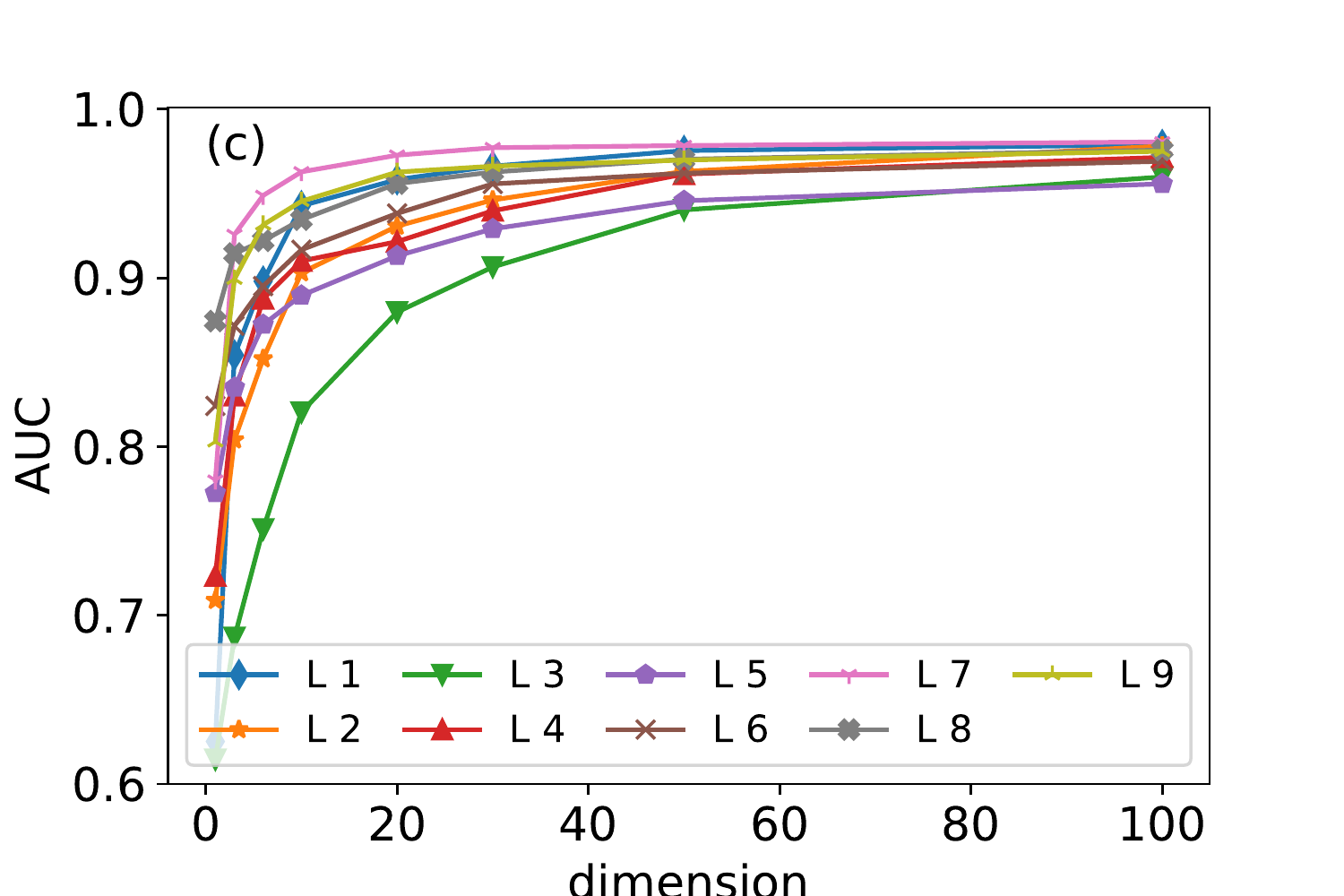}
\includegraphics[width=0.48\linewidth]{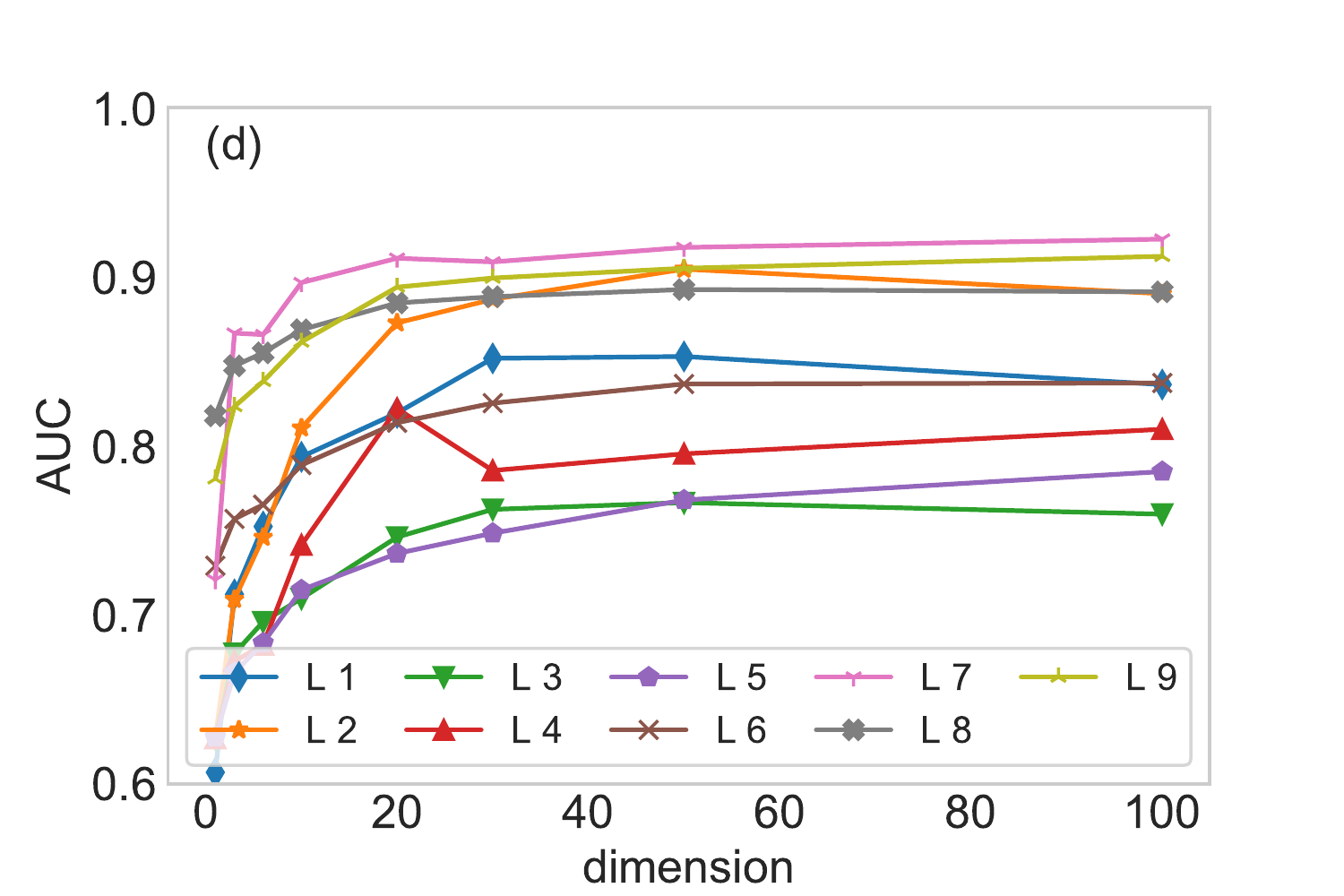}
\caption{\label{fig:fig4} AUC vs. the dimension of node vectors. \textcolor{blue}{Panel (a)} shows the MLE method on the first nine layers of the FAO network, where each layer is set as the target layer respectively. \textcolor{blue}{Panel (b)} shows the results of MAP method when the five layers with highest similarities are adopted to compute the parameters of the conjugate prior. \textcolor{blue}{Panel (c)} shows the results of the MLE method on the HVR network. \textcolor{blue}{Panel (d)} is the results of MAP method on the HVR network. Note that the results are averaged over 10 \textcolor{blue}{runs of} cross-validations.}
\end{figure}

\paragraph{The dimension of node vectors.} The dimension of node vectors is a critical factor for the MAP algorithm. In experiments, we find that there is no such number of communities \cite{louvain, guide, com1, girvan, clauset} that can maximize the estimation accuracy. For both the FAO network and HVR network, we remove 40\% of edges from the target layer to generate incomplete networks. The five layers with highest SimHash similarities are used to compute the parameters of the conjugate prior. The target layer is reconstructed through both the MLE algorithm and the MAP algorithm. In Fig. \ref{fig:fig4}, we can observe an increased estimation accuracy with respect to the increment of the number of dimensions, though the gain diminishes for dimensions greater than 40. In the following experiments, we use a dimension of 50 which balances running time and estimation accuracy. \textcolor{blue}{Apart from the dimension of node vectors, the estimation accuracy is also affected by the number of similar layers. The influence of the number of similar layers on the estimation accuracy is analyzed in appendix E.}

\begin{figure}[ht]
\includegraphics[width=0.98\linewidth]{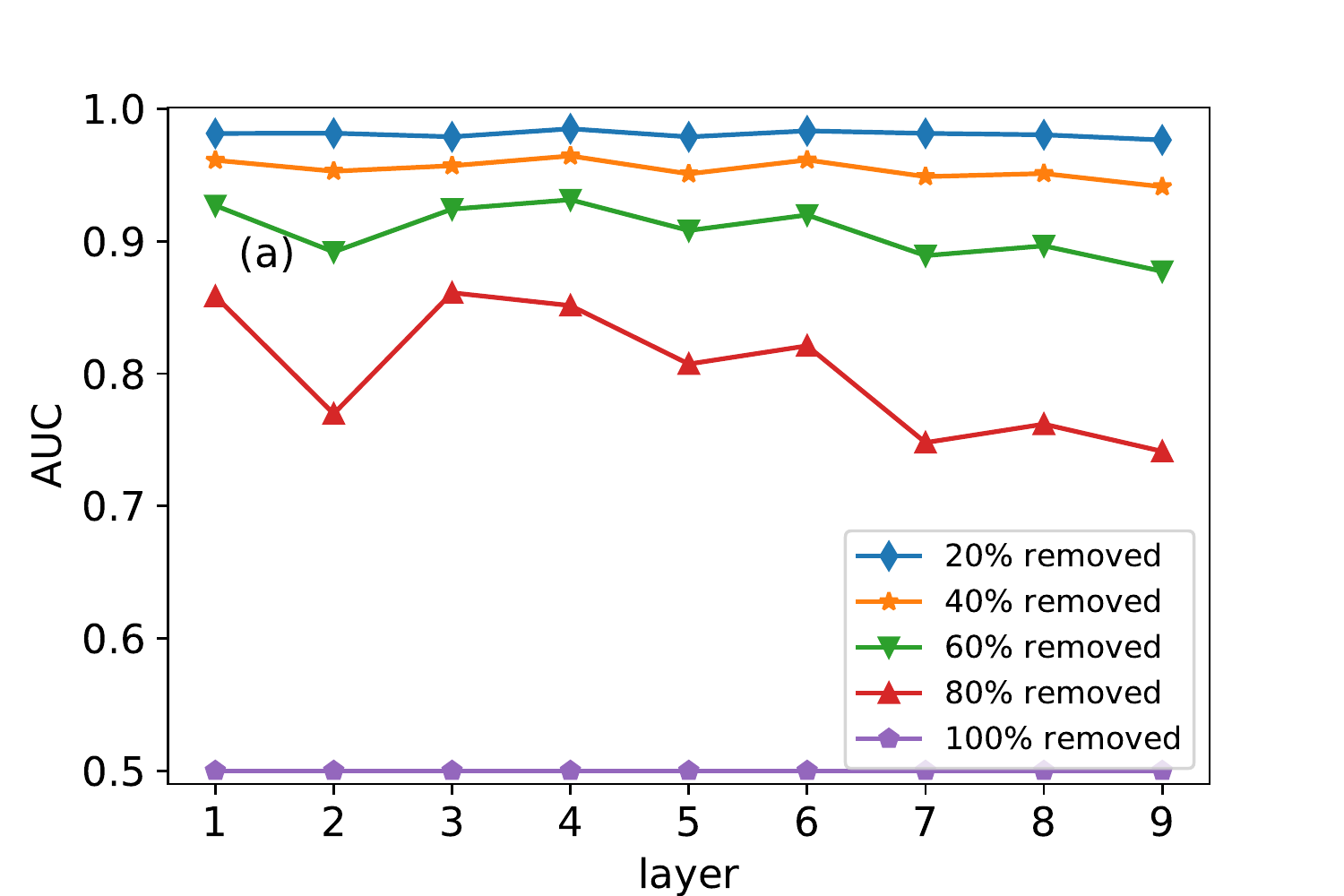}
\includegraphics[width=0.98\linewidth]{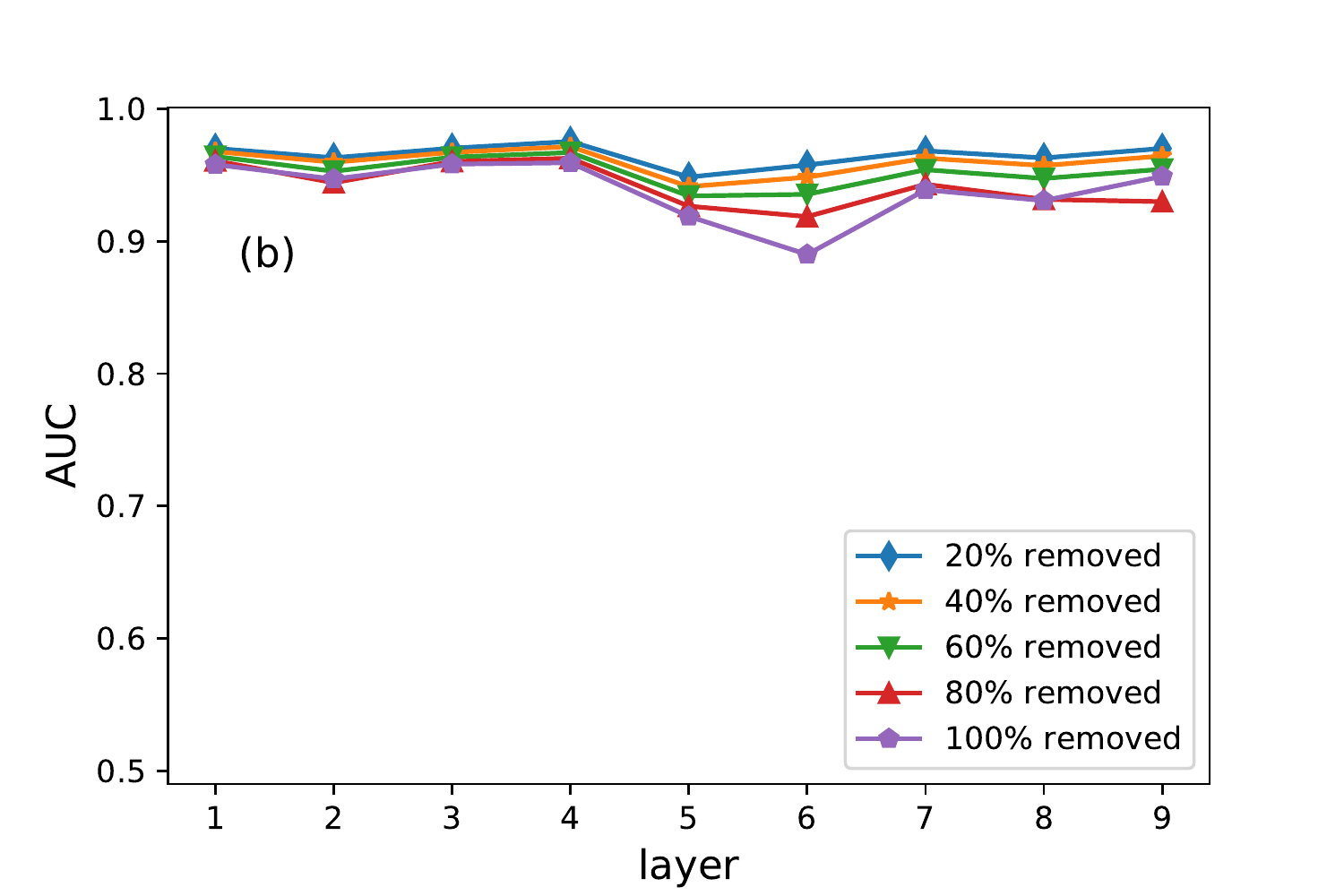}
\caption{\label{fig:fig5} Comparison of the MLE and MAP methods on the FAO network. \textcolor{blue}{Panel (a)} shows results of the MLE method. \textcolor{blue}{Panel (b)} shows the results of the MAP method. Results are averaged over 10 \textcolor{blue}{runs of} cross-validations.}
\end{figure}

\begin{figure}[ht]
\includegraphics[width=0.98\linewidth]{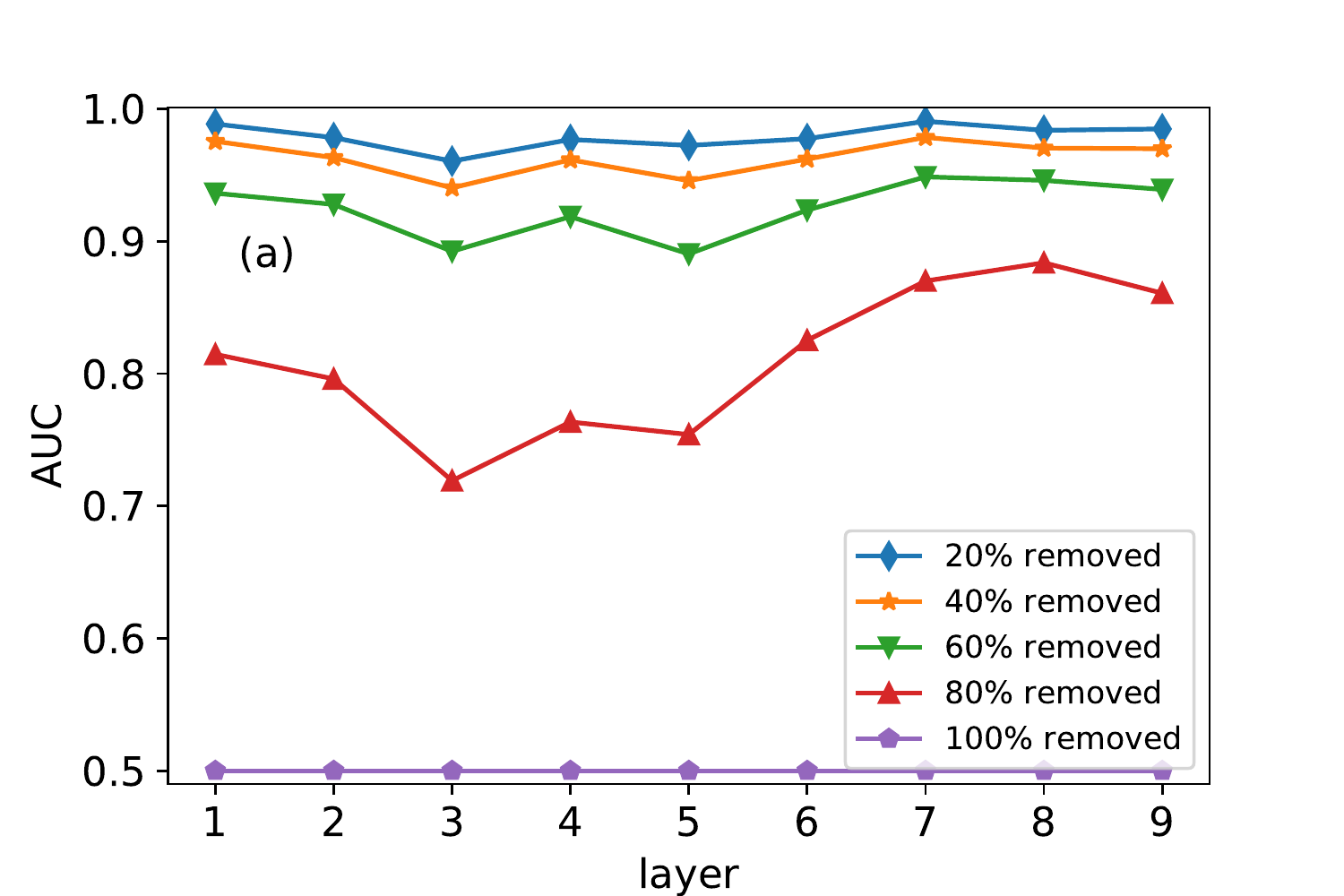}
\includegraphics[width=0.98\linewidth]{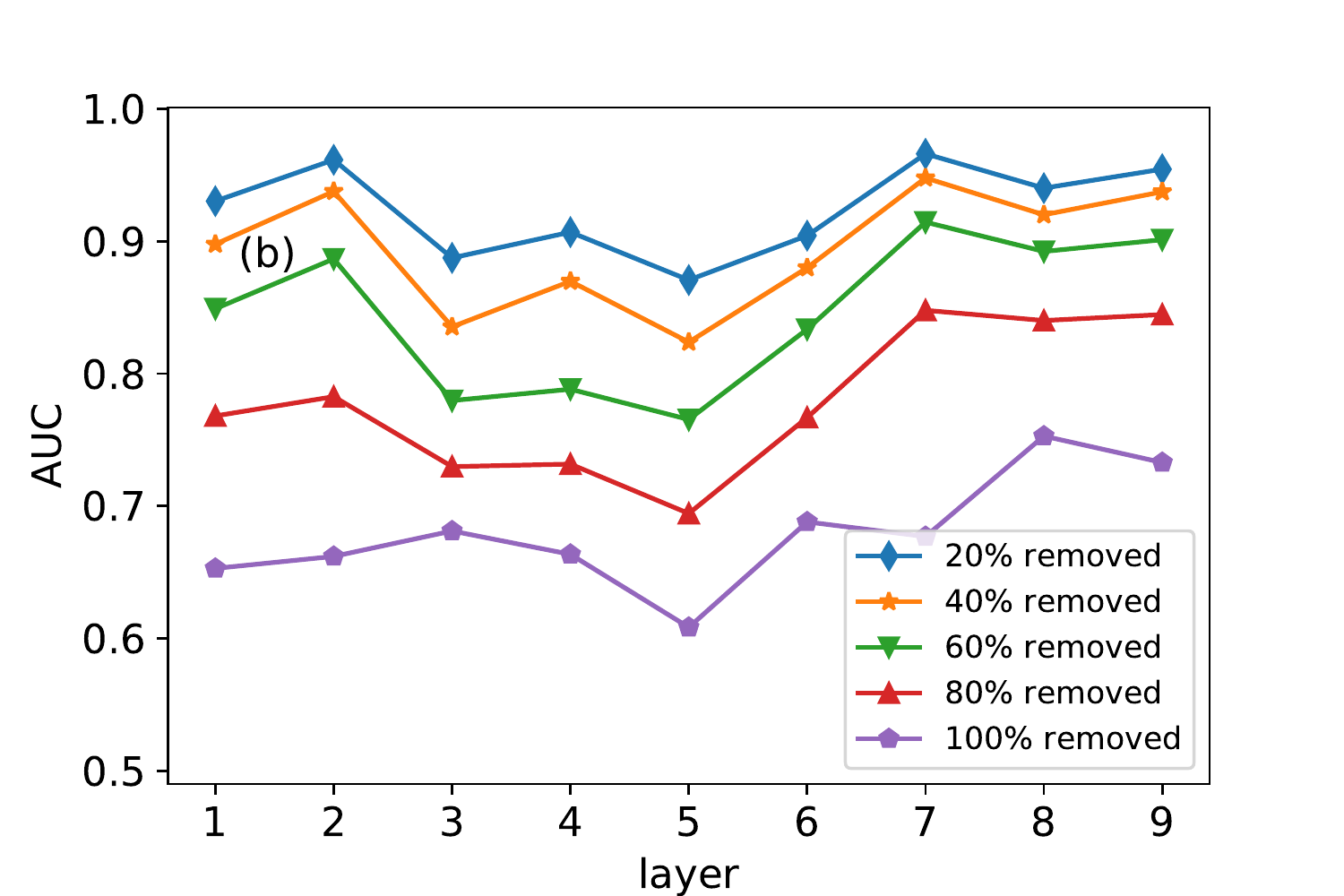}
\caption{\label{fig:fig6} Comparison of the MLE and MAP methods on the HVR network. \textcolor{blue}{Panel (a)} shows results of the MLE method. \textcolor{blue}{Panel (b)} shows the results of the MAP method. Results are averaged over 10 \textcolor{blue}{runs of} cross-validations.}
\end{figure}

\paragraph{Comparison between the MLE algorithm and the MAP algorithm.} Both the MLE algorithm and the MAP algorithm have their own benefits and disadvantages. The major difference between the MAP and MLE methods is that the MAP method can incorporate prior information (other similar layers), while the MLE relies on the available information of the target layer solely. The \textcolor{blue}{comparison between the MLE algorithm and the MAP algorithm} is performed on both the FAO network \textcolor{blue}{(first nine layers) and the HVR network. The two algorithms are implemented on incomplete layers, which are generated by randomly removing 20\%, 40\%, 60\%, 80\% and 100\% of edges from the two networks.} \textcolor{blue}{In Fig. \ref{fig:fig5}(a) and Fig. \ref{fig:fig6}(a), we can see that the estimation based on the MLE algorithm is significantly affected by the missing links. However, the robustness of the estimation is greatly improved after we adopt structurally similar layers to reconstruct the target layer, as shown in Fig. \ref{fig:fig5}(b). On the contrary, the estimation accuracy deteriorates if we adopt layers with heterogeneous structures to reconstruct the target layer, which is shown in Fig. \ref{fig:fig6}(b).}

Intuitively, the MLE method reconstructs the target layer through the known information \textcolor{blue}{of the target layer itself}. As a result, the estimation accuracy is \textcolor{blue}{related to} the available information, i.e., the percentage of known links. In the MAP algorithm, the estimation is not only affected by the known information of the target layer but also the similar layers. In real applications, if the percent of missing links is less than 20\%, it is recommended to use the MLE algorithm, \textcolor{blue}{since it provides more accurate results than the MAP algorithm. Conversely, the MAP algorithm is the better choice if researchers are unaware of the percentage of missing links.}

\section{Conclusion and future works}
In this paper, we present a novel MAP estimation-based algorithm for target layer reconstruction in multilayer networks. In multilayer networks, some layers are structurally similar; thus, we can take advantage of the similar layers to reconstruct the target layer. In section II, we first derive the \emph{maximum a posteriori} estimation for target layer reconstruction in multilayer network. Second, the eigenvector centrality-based SimHash algorithm is introduced to detect structurally similar layers. The SimHash algorithm compares network features, thus it is not affected by the missing links. Third, we introduce two scenarios to obtain the parameters of the Gamma conjugate prior. In the first case, where the target layer is partially known, the SimHash algorithm is adopted to detect structurally similar layers. In the second case, where the target layer is not known at all, functionally similar layers is used to compute the parameters of the conjugate prior. In section III, we first show that the eigenvector centrality-based SimHash algorithm is able to return consistent similarity levels for different percentages of missing links. Then, we show that the estimation accuracy can be improved by increasing the number of dimensions of node vectors, and the gain is diminishing for dimensions greater than 40 for the two networks. We find that with a great number of similar layers, we can obtain more consistent estimation results. Finally, the MLE method and MAP method are compared on two real networks. The experimental results suggest that if there are less than 20\% of missing links, the MLE method has better performance. However, if the percentage of missing links is 40\% or more, the MAP method returns results that are more consistent.

However, there are still some limitations to the MAP method we present. The first limitation is that the missing links in the target layer are required to be removed uniformly at random. The similarities are obtained through network feature comparison, which means the structure of the target layer needs to be maintained. Targeted removal of links will change the structure of the network. The second limitation concerns the unknown relation between the vector dimension and network size. In our numerical experiments, we test multiple dimensions and adopt a dimension of 50, which balances running time and estimation accuracy.

Recent results in \cite{eig1,eig2,eig3,eig4} present some methods to estimate the eigenvector centralities based on nodal data without requiring the network structure. Recall that SimHash algorithm is based on the eigenvector centrality obtained from the target layer. Therefore, estimating eigenvector centrality without constructing network is a good alternative for future work.

The MAP algorithm we present to reconstruct a target layer in a multilayer network shows promising results when we can identify the similarities between the target layer and other layers. The experimental results show that the estimations of our MAP method are less likely to be affected by missing links, which is not known in real applications. Therefore, our MAP method can be used to direct experiments, especially when there is no information about the target layer.

\section{Acknowledgements} 
This \textcolor{blue}{work has been} supported by the National Institutes of Health under Grant No. 1R01AI140760-01A1.

\appendix
\counterwithin{figure}{section}
\section{tensor factorization} 
In \cite{newman1, newman2, bacco}, nodes are factorized by membership vectors. The dimension of membership vectors is interpreted as the number of overlapping communities. In addition, each edge is computed through the tensor product of the membership vectors. The number of communities is obtained by maximizing the likelihood. Vectorization is also used in some natural language processing algorithms \cite{bert, glove, peters, word2vec1, word2vec2} in which words are embedded as vectors to preserve their relations to contexts. However, the dimension of word vectors does not have any semantic meaning, rather the choice of the vector dimension is first affected by the data set size. According to \cite{glove, word2vec2}, larger dimensions can improve model accuracy, but the gain diminishes for vectors larger than 200 dimensions. The choice of dimension is also related to the available resources, and it is better to reduce the dimension as long as the choice does not affect the estimation accuracy substantially. \textcolor{blue}{In fact, the tensor factorization can be simplified to dot product, we prove this point as follows.}

We assume the number of edges between any two nodes is the tensor product of two node vectors and the control matrix, i.e., $E_{ij} = \sum s_{ik} t_{jl} w_{kl}$. $w_{kl}$ ($k$ is not equal to $l$) is the parameter that controls the edges from any source node $i$ (in community $k$) to any target node $j$ (in community $l$). Then, we have
\begin{align}
\label{eq:a1}
    E_{ij}&=\sum s_{ik} t_{jl} w_{kl}\nonumber\\
    &=s_{ik} t_{jk} w_{kk}+s_{ik} t_{jl} w_{kl}+s_{il} t_{jl} w_{ll}\nonumber\\
    &+ (\sum s_{im} t_{jn} w_{mn}- s_{ik} t_{jk} w_{kk}-s_{ik} t_{jl} w_{kl}-s_{il} t_{jl} w_{ll}).
\end{align}

In equation \ref{eq:a1}, we denote the sum of the first three terms as $E_{ij}^{kl}$. Then, we can incorporate $w_{kk}$ into $s_{ik}$, and denote it as $s_{ik}^\prime=s_{ik} w_{kk}$. Similarly, $s_{il}^\prime=s_{il} w_{ll}$. Therefore, we have 
\begin{align}
\label{eq:a2}
    E_{ij}^{kl} &=s_{ik}^\prime t_{jk} + s_{ik}^\prime t_{jl} \frac{w_{kl}}{w_{kk}} + s_{il}^\prime t_{jl}\nonumber\\
    &=s_{ik}^\prime t_{jk}+ (s_{ik}^\prime \frac{w_{kl}}{w_{kk}} + s_{il}^\prime) t_{jl}.
\end{align}

Since nodes between different communities are loosely connected, we have $w_{kl}<w_{kk}$, and we assume $\epsilon=w_{kl}/w_{kk}$ . Then, equation \ref{eq:a2} can be written as 
\begin{align}
\label{eq:a3}
    E_{ij}^{kl} &=s_{ik}^\prime t_{jk}+ (\epsilon s_{ik}^\prime + s_{il}^\prime) t_{jl} \nonumber\\
    &=s_{ik}^\prime t_{jk}+ s_{il}^{\prime\prime} t_{jl},
\end{align}
where $s_{il}^{\prime\prime}=\epsilon s_{ik}^\prime + s_{il}^\prime$. For every inter-community edge originating from node $i$ in community $k$, we can incorporate it into community $l$ by incrementing $\epsilon s_{ik}$ to $s_{il}^\prime$. \textcolor{blue}{Edges can be factorized through the dot product of node vectors.  Therefore, the tensor factorization is equivalent to the dot product.}

\section{Estimation based on the Pearson correlation}
\textcolor{blue}{With the SimHash algorithm, the weighted digest of each layer can be obtained. The similarity between any two layers can be measured by computing the Pearson correlation of the weighted digests. In Fig. \ref{fig:figAB}, we show the estimation results based on the Pearson correlation coefficient. We observe that the estimation is as consistent as the results based on the Hamming distance.} Therefore, the correlation can be an alternative for the Hamming distance.

\textcolor{blue}{\begin{figure}[ht]
\includegraphics[width=0.98\linewidth]{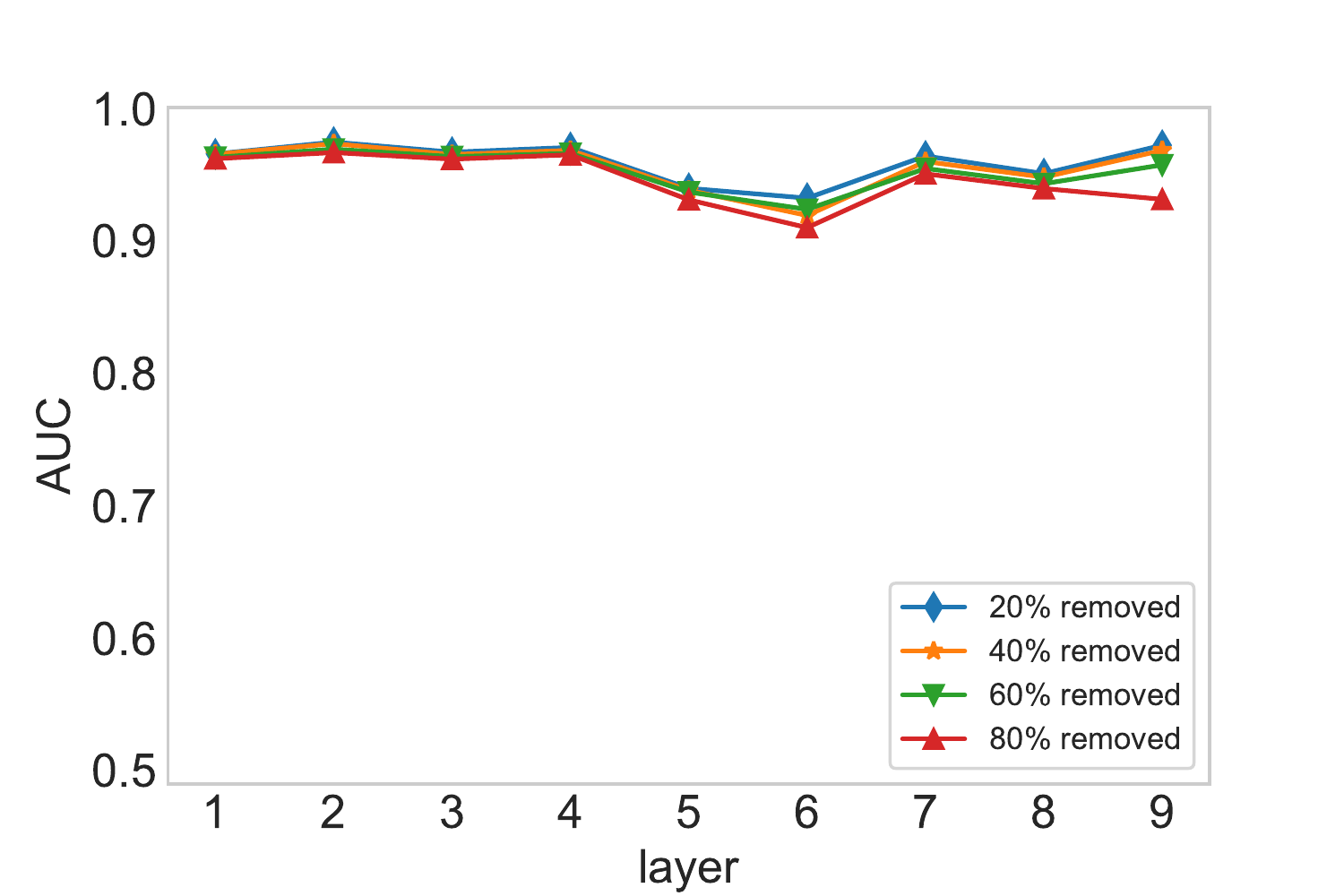}
\caption{\label{fig:figAB} \textcolor{blue}{The estimation results based on the Pearson correlation of the weighted digest.}}
\end{figure}}
 
\section{The choice of the number of bits}
\begin{figure}[ht]
\includegraphics[width=0.48\linewidth]{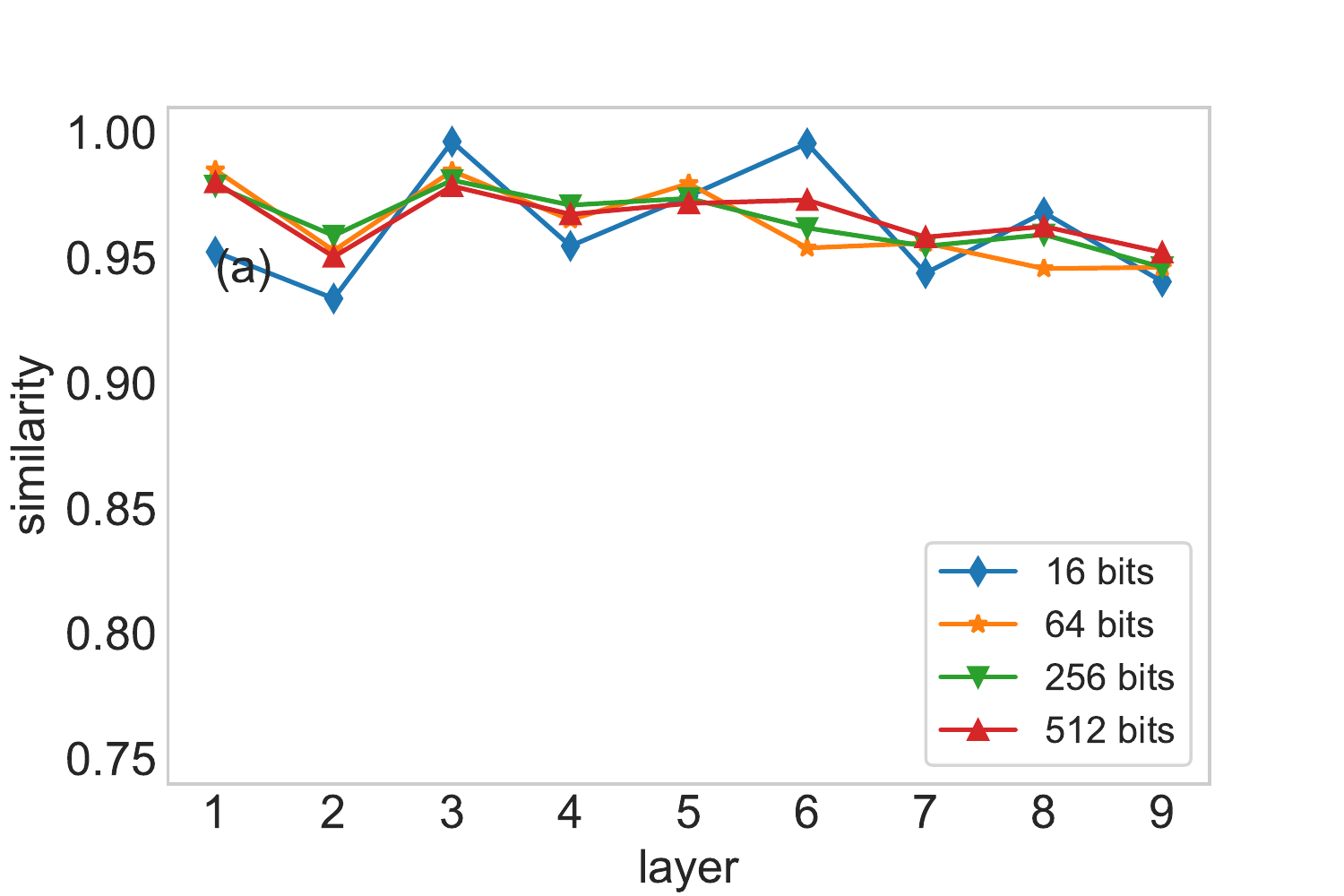}
\includegraphics[width=0.48\linewidth]{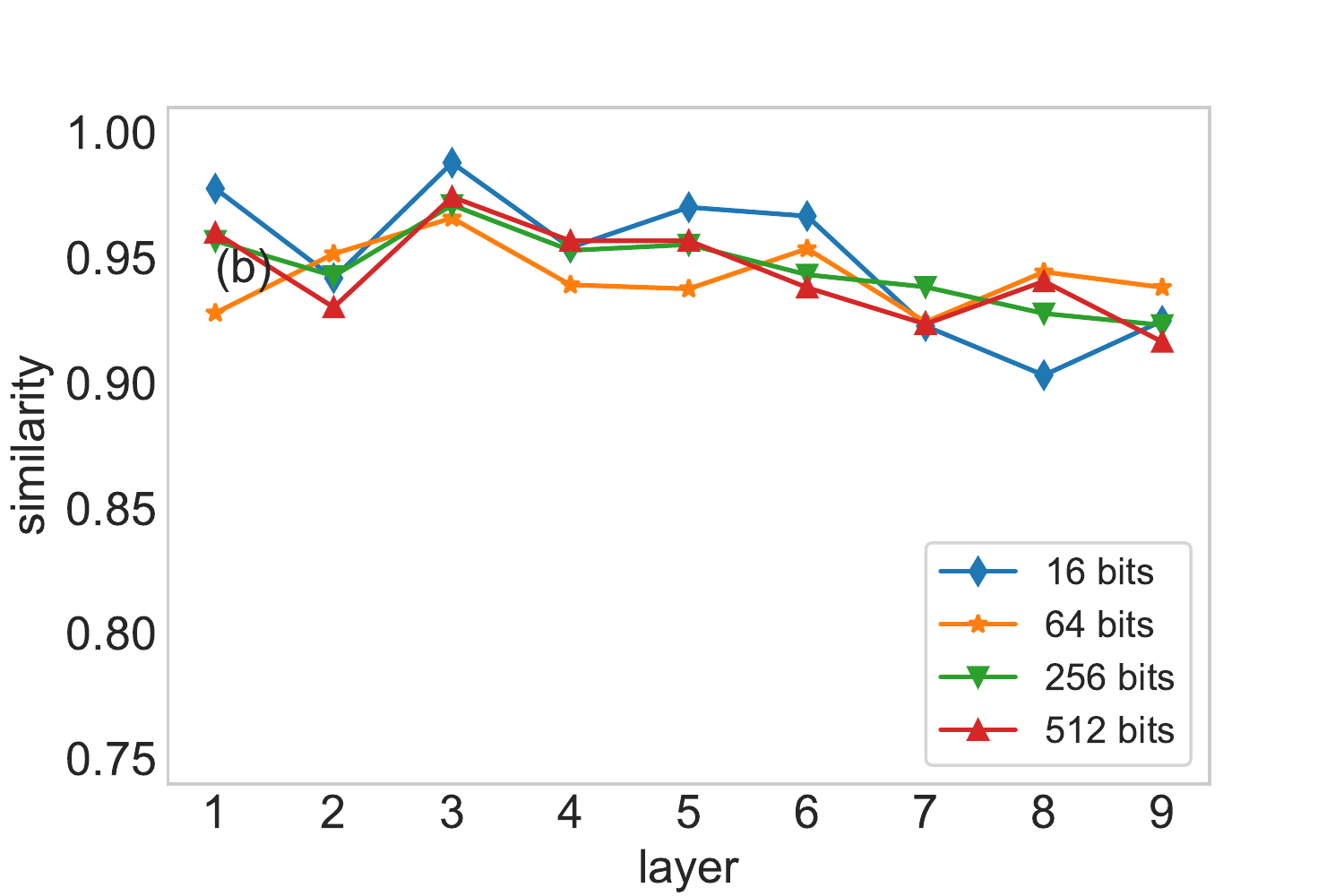}
\includegraphics[width=0.48\linewidth]{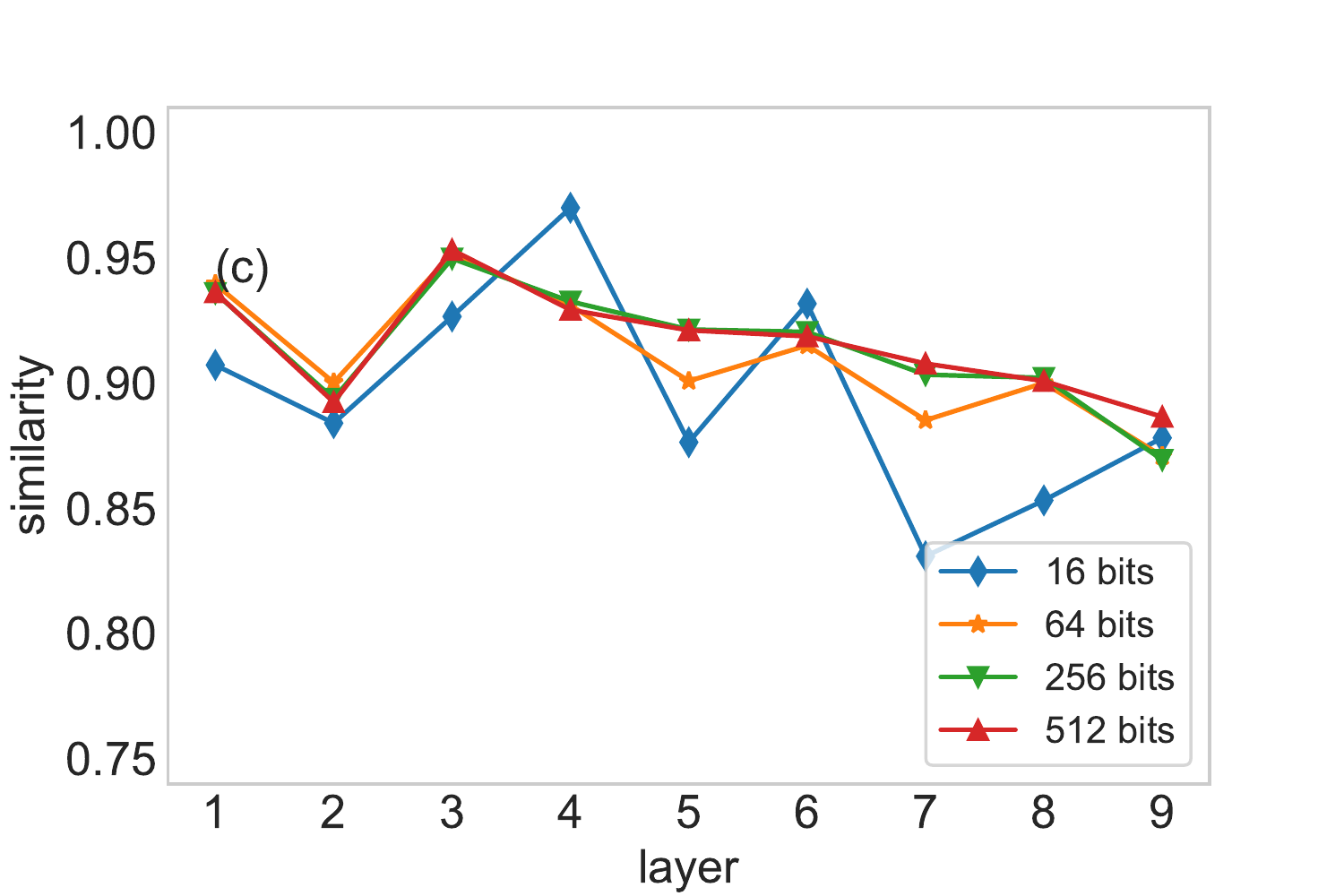}
\includegraphics[width=0.48\linewidth]{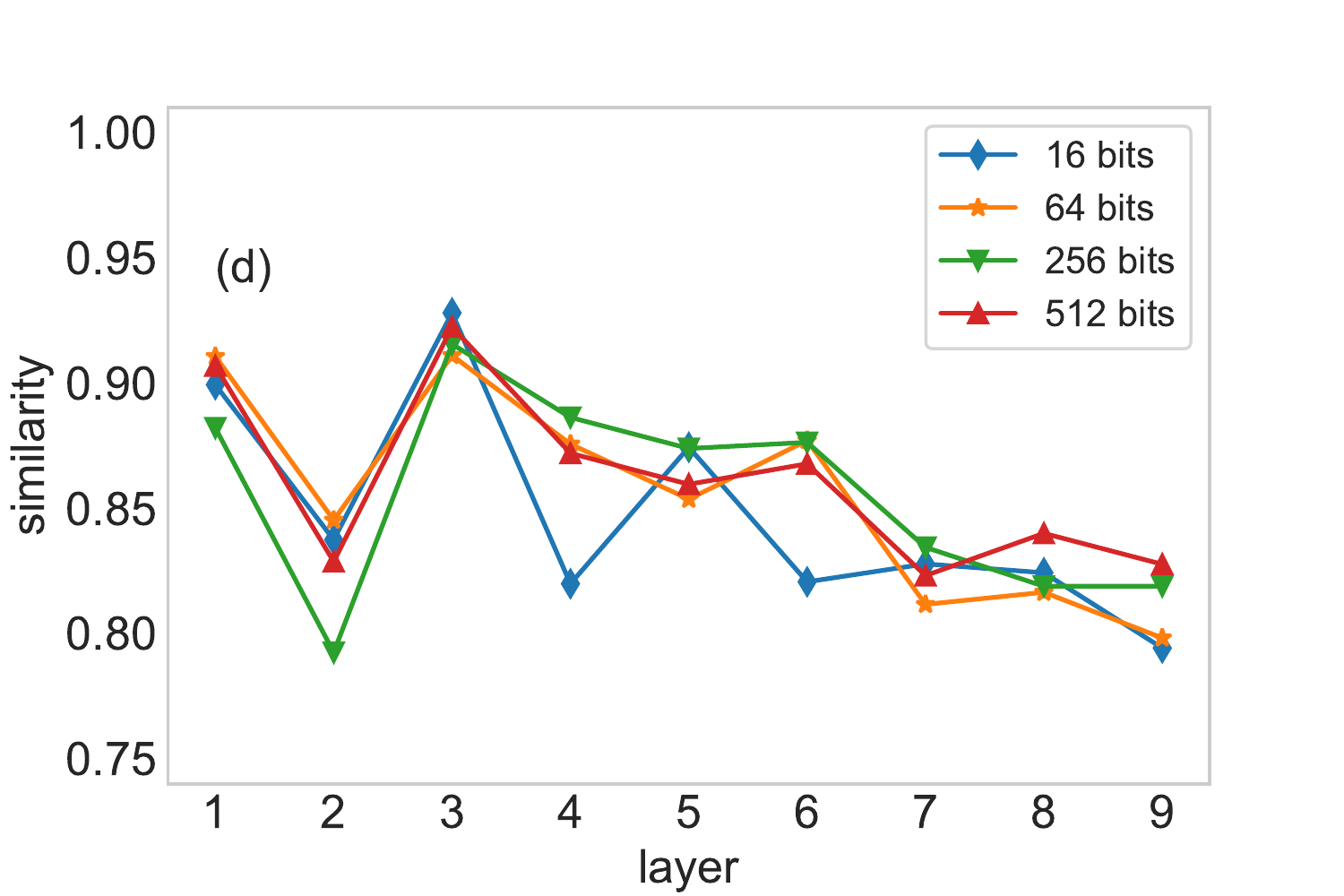}
\caption{\label{fig:figAC1} Comparison of varying the number of digits (based on the Hamming distance). The four panels show the results based on 20\% (a), 40\% (b), 60\% (c), and 80\% (d) of edges randomly removed.}
\end{figure}

The choice of $\phi$ affects the resolution and the consistency of the similarity. If the number of tokens is large, it is recommended to employ large $\phi$. In our work, the sizes of the two networks are not too large, so we adopt $\phi =512$ bits. In Fig. \ref{fig:figAC1}, we validate this choice on the FAO network. We set each of the first nine layers of the FAO network as the target network (original network), and randomly remove 20\%, 40\%, 60\% and 80\% of edges from the target network to generate incomplete networks. We then compare the incomplete networks with the original network. In the experiments, we adopt $\phi=16,\ 64,\ 256$, and $512$ digits, respectively. We can observe that the similarities obtained with 256 digits are close to that obtained with 512 digits, while the similarities obtained with 16 and 64 digits vary remarkably.

\section{Comparison between the MLE algorithm and entropy-based approaches} 
In section III B, we compared the MAP algorithm with the MLE algorithm. In fact, the entropy-based approaches are equivalent to the MLE algorithm. In the following, we prove that the MLE algorithm is equivalent to the entropy-based approaches.

Given a network $G$, we assume the number of nodes is fixed, and the edge weights are variables. Thus, we can define the entropy of a network in terms of edge weights as
\begin{align}
\label{eq:b1}
H(G)&=-\sum_G \sum_u p^U(u) \log p^U(u) \nonumber\\
&=\sum_G \sum_u p^U(u) \log \frac{1}{p^U(u)},
\end{align}
where $U$ is any edge in the network $G$, $u$ is the weight of edge $U$. $p^U (u)$ is the precise probability of edge $U$ with weight $u$. We assume for any edge $U$, it can take $n_U$ values. If the $n_U$ values follow uniform distribution, then we have $p^U (u)=\frac{1}{n_U}$. In this case, the network has the largest entropy, which means the network is totally uncertain. On the other hand, for any edge $U$, if there is a value $u_s$ with $p^U (u_s )=1$, the entropy of the network will be zero, which indicates all the edge weights in the network are known.

In our problem, the goal is to find a model to reconstruct the target layer, thus, we need a set of parameters to describe the model. If the model is close to the true network, the entropy of the model is minimized. Hence, the problem is transformed to finding the parameters of a model with minimum entropy, i.e., reducing uncertainty.

Equation \ref{eq:b1} can be written in expectation form
\begin{align}
\label{eq:b2}
   H(p)= \sum_G E_{U\sim p^U(u)} (\log \frac{1}{p^U(u)}).
\end{align}

Consider the Kullback-Leibler divergence (KL divergence). We assume $\theta$ is the parameter set that can describe the model, the probability of edge $U$ with weight $u$ is $q^U (u;\theta)$. Thus, the relative entropy is
\begin{align}
\label{eq:b3}
   D_{kl} (p||q)&=\sum_G\sum_u p^U (u) \log \frac{p^U(u)}{q^U (u;\theta)}\nonumber\\
   &=\sum_G E_{U\sim p^U(u)} (\log \frac{p^U(u)}{q^U (u;\theta)})\nonumber\\
   &=\sum_G E_{U\sim p^U(u)} (\log \frac{1}{q^U (u;\theta)})\nonumber\\
   &- \sum_G E_{U\sim p^U(u)} (\log \frac{1}{p^U (u)}).
\end{align}

The second term in the right-hand side is equation \ref{eq:b2} and the first term in the right-hand side is the cross entropy, which is
\begin{align}
\label{eq:b4}
H(p,q)=\sum_G E_{U\sim p^U (u)} \log \frac {1}{q^U (u;\theta)}.
\end{align}

Recall that our problem is to reconstruct the target layer and equation \ref{eq:b2} is the entropy of the target layer. Equation \ref{eq:b2} is determined by the data (network) solely. Thus, the problem can be simplified to minimizing the cross entropy, i.e., minimizing equation \ref{eq:b4}.

Then, we consider the MLE method. The MLE algorithm is to find the parameter set $\theta$ that most likely fit a given set of data (network) $D$. We skip some intermediate steps and use the logarithm form directly as following
\begin{align}
\label{eq:b5}
\theta &= argmax_\theta P(G\ |\ \theta)\nonumber\\
&=argmax_\theta \sum_G \sum_u \log p^U(u;\theta).
\end{align}

$p^U (u;\theta)$ is the model with parameter set $\theta$ to describe the true data $D$. We assume that edge $U$ can take $n_U$ values and the $n_U$ values follow the uniform distribution, then we have $p_{MLE}^U (u)=\frac{1}{n_U}$. Introducing $p_{MLE}^U$ to equation \ref{eq:b5} does not change the results, we have
\begin{align}
\label{eq:b6}
\theta &= argmax_\theta \sum_G \sum_u \frac{1}{n_U}\log p^U(u;\theta) \nonumber\\
&=argmax_\theta \sum_G E_{u\sim p_{MLE}^U(u)} (\log p^U(u;\theta)).
\end{align}

Our goal is still to find a set of parameters that can reconstruct the layer. Then, we can generalize this problem by replacing $p_{MLE}^U(u)$ with $p^U (u)$, and replacing $p^U(u;\theta)$ with $q^U (u;\theta)$. The replacements can be regarded as taking real data (network) into equation \ref{eq:b6}. We have
\begin{align}
\label{eq:b6}
\theta &= argmax_\theta \sum_G E_{u\sim p^U(u)}\log q^U(u;\theta) \nonumber\\
&=argmax_\theta \sum_G -E_{u\sim p^U(u)}\log q^U(u;\theta) \nonumber\\
&=argmax_\theta \sum_G E_{u\sim p^U(u)}\log \frac{1}{q^U(u;\theta)}. 
\end{align}

Thus, the MLE algorithm is equivalent to minimizing the cross entropy.

\section{The number of similar layers on the estimation accuracy}

\begin{figure}[t!]
\includegraphics[width=0.48\linewidth]{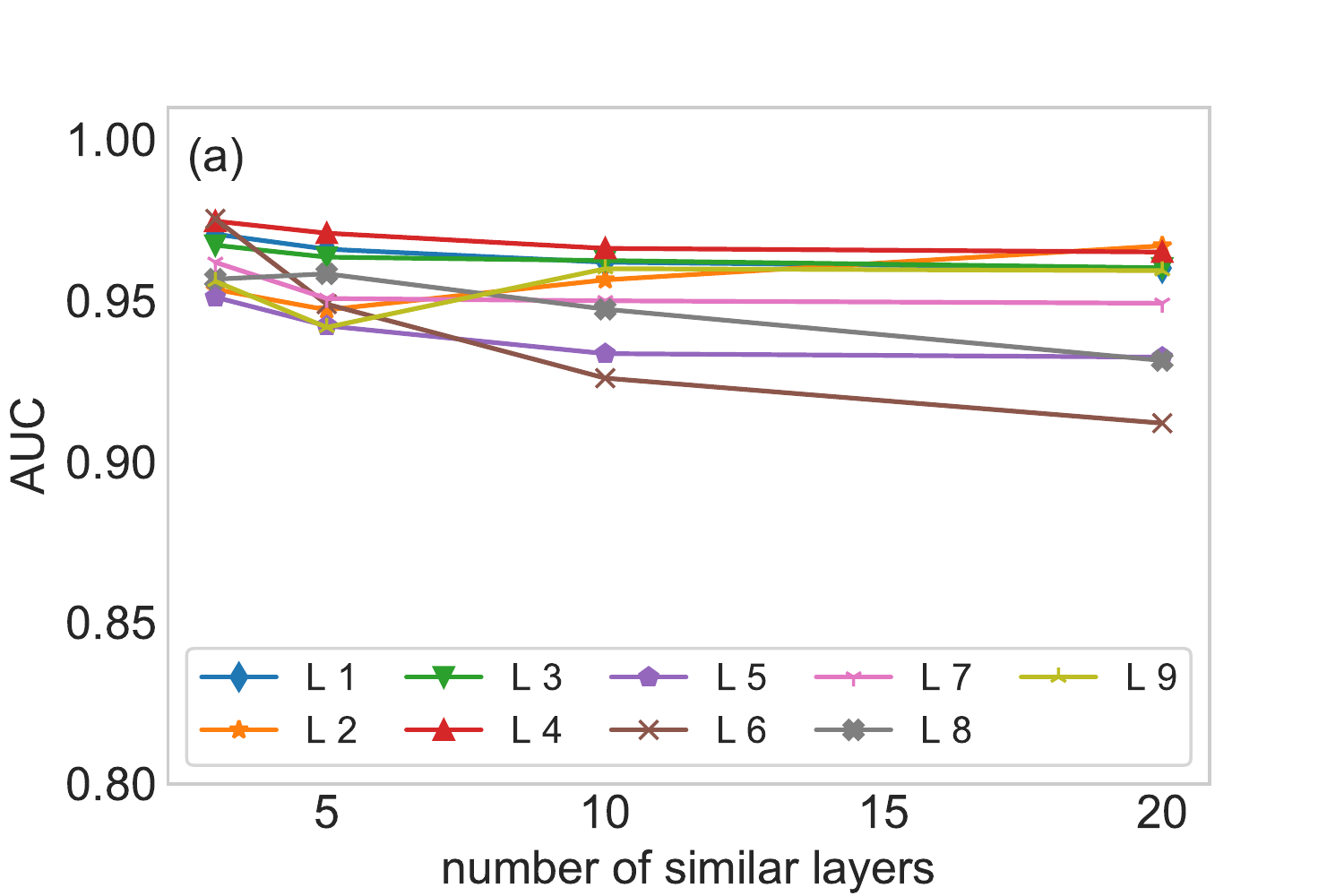}
\includegraphics[width=0.48\linewidth]{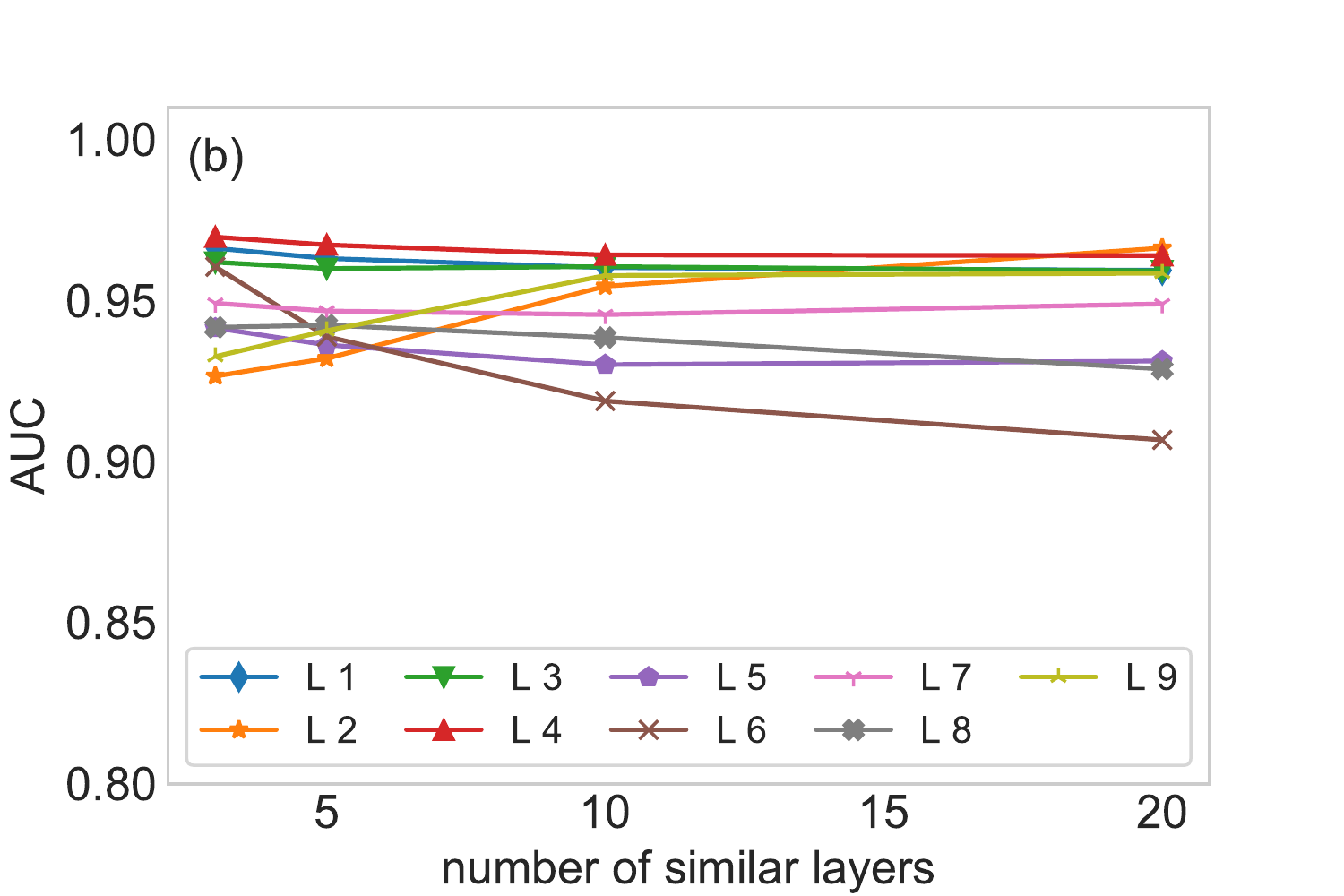}
\includegraphics[width=0.48\linewidth]{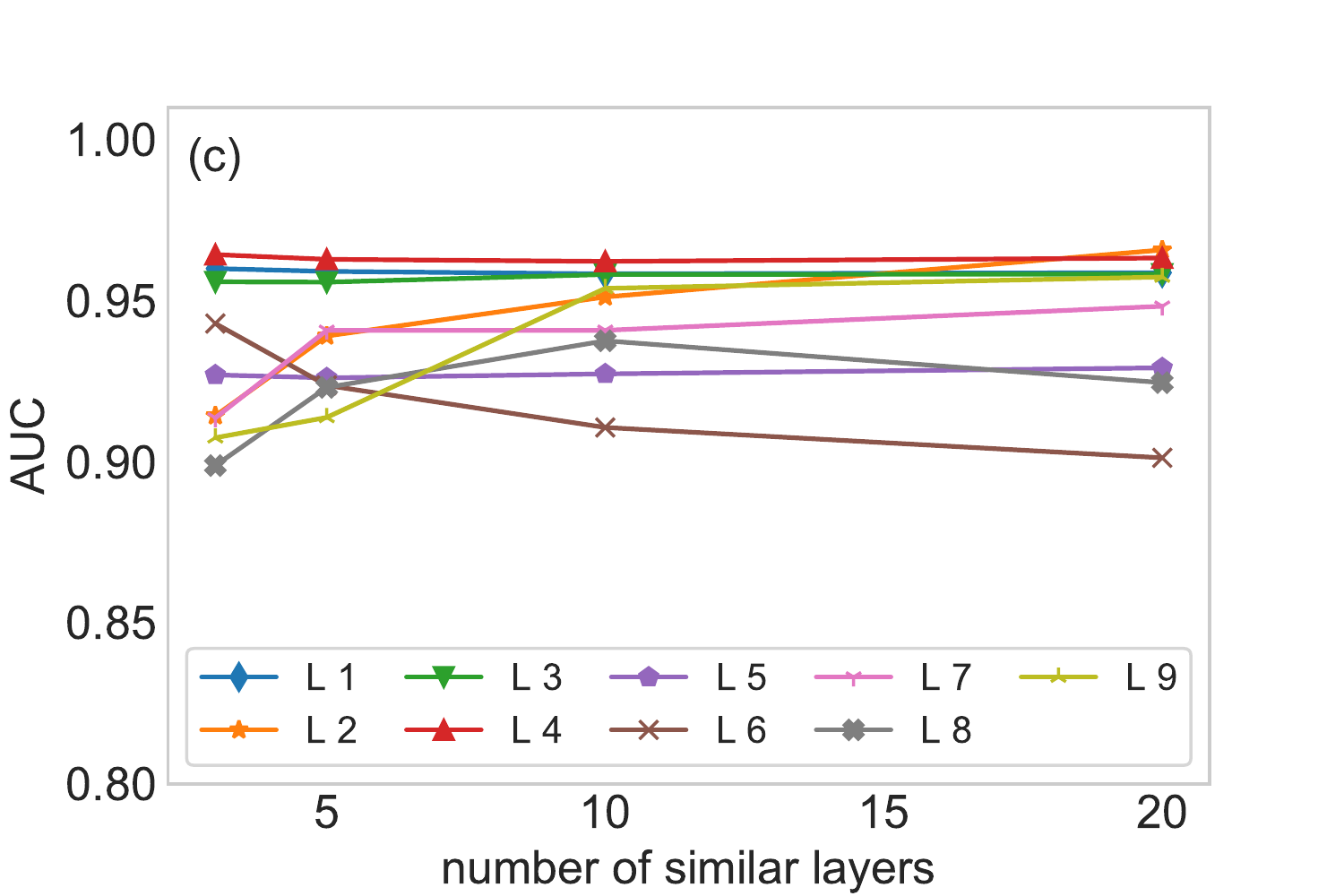}
\includegraphics[width=0.48\linewidth]{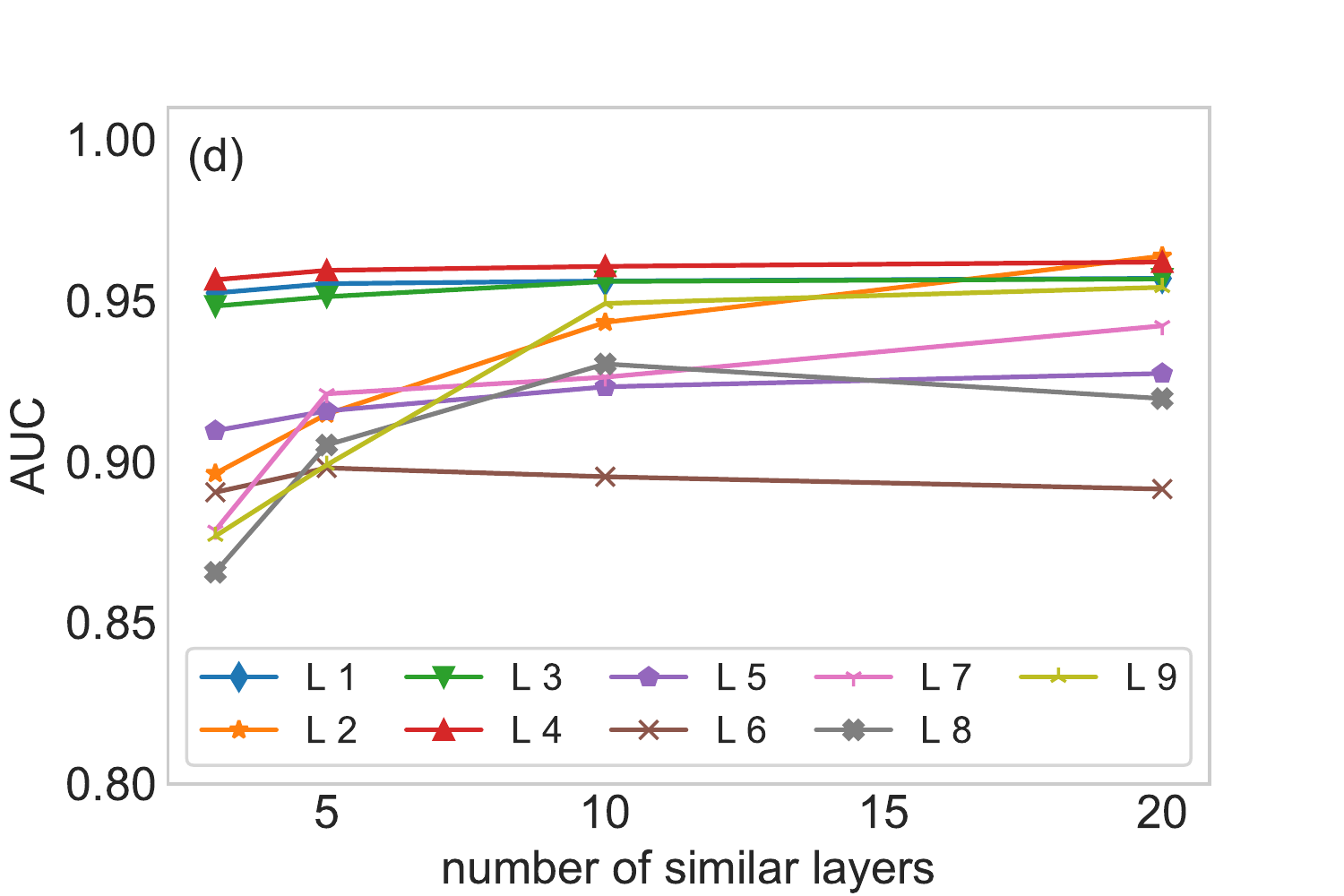}
\caption{\label{fig:figAE1} \textcolor{blue}{The number of similar layers on the estimation results. 20\% (a), 40\% (b), 60\% (c), and 80\% (d) of edges are randomly removed from the target layer. Different numbers of} similar layers are used to compute the parameters of the conjugate prior. The experiments are conducted on the first nine layers of the FAO network. Each AUC is averaged over 10 \textcolor{blue}{runs of} cross-validations.}
\end{figure}

If more similar layers are employed to compute the parameters of the conjugate prior, the influence of similar layers on the reconstruction will be promoted. \textcolor{blue}{Consequently}, the influence of the known part of the target layer will be down weighted. On the contrary, if we trust the known part of the target layer, we can employ less similar layers to compute the parameters of the conjugate prior. The experimental results are shown in Fig. \ref{fig:figAE1}, \textcolor{blue}{top 3, 5, 10, and 20 similar layers are adopted to compute the parameters of the conjugate prior. We observe that the estimation based on 10 and 20 similar layers are robust with respect to the missing links, while the estimation based on three and five similar layers are influenced by the missing links significantly.}

\newpage


\end{document}